\documentclass[a4paper,11pt]{article}
\usepackage{graphicx}
\usepackage{bm}
\usepackage{subfigure}
\usepackage{url}
\usepackage[svgnames]{xcolor}
\usepackage[ddmmyy,24hr]{datetime}
\usepackage{bigdelim}
\usepackage{booktabs}
\usepackage{dcolumn}
\usepackage{multirow}
\usepackage{subfigure}
\usepackage{cancel}
\usepackage{stackrel}
\usepackage{paralist}
\usepackage{xspace}
\usepackage{slashed}
\usepackage{enumerate}
\usepackage{enumitem}
\usepackage{soul}
\usepackage{diagbox}
\usepackage{tabularx}
\usepackage{graphicx}
\usepackage{jheppub}

\newcommand{\nua}[1]{\ensuremath{\rlap{\kern-2.5pt\ensuremath{\overset{\scriptscriptstyle(-)}{\phantom{\nu}}}}{\ensuremath{{\nu}_{#1}}}}}

\title{Neutrinoless double beta decay in the minimal type-I seesaw model: mass-dependent nuclear matrix element, current limits and future sensitivities}

\author[a,b]{Dong-Liang~Fang}
\author[c,d]{Yu-Feng~Li}
\author[c]{Yi-Yu~Zhang}
\author[a,b]{Jing-Yu~Zhu}
\affiliation[a]{Institute of Modern Physics, Chinese Academy of sciences, Lanzhou, 730000, China}
\affiliation[b]{School of Nuclear Science and Technology, University of Chinese Academy of Sciences, Beijing 100049, China}
\affiliation[c]{Institute of High Energy Physics, Chinese Academy of Sciences, Beijing 100049, China}
\affiliation[d]{School of Physical Sciences, University of Chinese Academy of Sciences, Beijing 100049, China}

\emailAdd{dlfang@impcas.ac.cn, liyufeng@ihep.ac.cn,
zhangyiyu@ihep.ac.cn,
zhujingyu@impcas.ac.cn}

\abstract{In this work we discuss the neutrino mass dependent nuclear matrix element (NME) of the neutrinoless double beta decay process and derive the limit on the parameter space of the minimal Type-I seesaw model from the current available experimental data as well as the future sensitivities from the next-generation experiments. 
Both the explicit many-body calculations and naive extrapolations of the mass dependent NME are employed in the current work. The uncertainties of the theoretical nuclear structure models are taken into account. By combining the latest experimental data from  $^{76}$Ge-based experiments, GERDA and MAJORANA, the $^{130}$Te-based experiment, CUORE and the $^{136}$Xe-based experiments, KamLAND-Zen and EXO-200, the bounds on the parameter space of the minimal Type-I seesaw model are obtained and compared with the limits from other experimental probes. Sensitivities for future experiments utilizing 
$^{76}$Ge-based (LEGEND-1000), $^{82}$Se-based (SuperNEMO), $^{130}$Te based (SNO+II) and $^{136}$Xe-based (nEXO), with a ten-year exposure, are also derived.}	

\begin{document}	
\maketitle
\section{Introduction}\label{sec:introduction}
Neutrinos, as the building blocks of the Standard Model (SM), are the most mysterious fermion of the fundamental particles. The recent groundbreaking discovery of neutrino oscillations \cite{Kajita:2016cak,McDonald:2016ixn} has revealed that neutrinos are massive, but their masses are significantly smaller than those of other charged fermions.
The nature of electric neutrality allows for the possibility of neutrinos being Majorana fermions, which opens up the potential for explaining the smallness of neutrino masses via the seesaw mechanisms~\cite{Xing:2020ijf}. There are three distinct types of seesaw mechanisms according to the different SM group representations of the right-handed neutrinos (RHNs) and/or Higgs particles~\cite{Minkowski:1977sc, Yanagida:1979as,Gell-Mann:1979vob, Glashow:1979nm, Mohapatra:1979ia,Konetschny:1977bn, Magg:1980ut, Schechter:1980gr, Cheng:1980qt, Mohapatra:1980yp, Lazarides:1980nt,Foot:1988aq}. 
The most popular realization is the canonical type-I seesaw mechanism~\cite{Minkowski:1977sc, Yanagida:1979as,Gell-Mann:1979vob, Glashow:1979nm, Mohapatra:1979ia}, which introduces at least two SM-singlet Majorana RHNs~\cite{Frampton:2002qc,Xing:2020ald,Guo:2006qa,Zhao:2022trb}. The active neutrino masses are obtained by diagnolizing the entire neutrino mass matrices and their smallness is attributed to the heavy masses of Majorana RHNs.
The models of seesaw mechanisms have far-reaching consequences, particularly in terms of lepton number violation processes in the field of particle and nuclear physics. These include the neutrinoless double beta decay ($0\nu\beta\beta$) process ~\cite{Bilenky:2014uka,Dolinski:2019nrj,Agostini:2022zub,Engel:2016xgb,Yao:2021wst}, 
the lepton number violating decays of $K$ or charmed mesons~\cite{Chun:2019nwi,Isidori:2014rba} and the search for lepton number violating signals at colliders~\cite{Cai:2017mow}. These phenomena provide valuable insights into the nature of neutrino masses and new physics beyond the SM.

The double beta decay is an extremely rare process in the universe, governed by the strong nuclear pairing force that enhances the stability of even-even nuclei compared to odd-odd nuclei~\cite{Engel:2016xgb,Yao:2021wst}. 
A unique mode of the double beta decay without the emission of neutrinos, known as the $0\nu\beta\beta$, is of utmost significance since it holds the key to unraveling the mystery of whether neutrinos are their own antiparticles, and thus being a direct probe of the Majorana nature of neutrinos.
Experimental searches for 
the $0\nu\beta\beta$ have been actively conducted worldwide~\cite{Bilenky:2014uka,Dolinski:2019nrj,Agostini:2022zub}, with the aim of detecting this extremely rare process and shedding light on the fundamental properties of neutrinos. The results of the $0\nu\beta\beta$ searches have been reported in various experiments using different isotopes, including $^{136}$Xe-based experiments KamLAND-Zen~\cite{KamLAND-Zen:2012mmx,KamLAND-Zen:2016pfg,KamLAND-Zen:2022tow} and EXO-200~\cite{EXO-200:2012pdt,EXO-200:2014ofj,EXO:2017poz,EXO-200:2019rkq}, $^{76}$Ge-based experiments GERDA~\cite{GERDA:2013vls,Agostini:2017iyd,GERDA:2018pmc,GERDA:2019ivs,GERDA:2020xhi} and MAJORANA~\cite{Majorana:2017csj,Majorana:2019nbd,Majorana:2022udl}, $^{130}$Te-based experiment, CUORE~\cite{CUORE:2017tlq,CUORE:2019yfd,CUORE:2021mvw}, and experiments using other isotopes such as $^{48}$Ca~\cite{Brudanin:2000in,Ogawa:2004fy,NEMO-3:2016mvr}, $^{82}$Se~\cite{Arnold:2018tmo,NEMO:2005xxi,CUPID:2019gpc}, $^{100}$Mo~\cite{Ejiri:2001fx,NEMO:2005xxi,NEMO-3:2013pwo,NEMO-3:2015jgm,Alenkov:2019jis,CUPID:2020aow}, $^{116}$Cd~\cite{NEMO-3:2016zfx}, and $^{150}$Nd~\cite{NEMO:2008kpp,NEMO-3:2016qxo}.
The most stringent lower limits on the $0\nu\beta\beta$ half-lives of $^{76}$Ge, $^{135}$Xe, and $^{130}$Te are derived at the levels of approximately $10^{25}$ to $10^{26}$ years, 
while the limits for the other isotopes are at the levels of $10^{24}$ years or below. By taking account of the large spread of the nuclear matrix elements (NMEs), these lower limits on half-lives can be converted to corresponding upper limits on the effective light Majorana neutrino mass, ranging from around 50 meV to several eV~\cite{Bilenky:2014uka,Dolinski:2019nrj,Agostini:2022zub,Pompa:2023jxc}.

In the framework of the Type-I seesaw model, the $0\nu\beta\beta$ process is 
directly affected 
by the masses of light and heavy neutrinos within the neutrino propagator, commonly known as the neutrino mass mechanism. Apart from the common phase space factor, the half-lives of the $0\nu\beta\beta$ depend on both the lepton part of the decay rate, denoted as the effective neutrino mass, and the nuclear part, represented by the NMEs.
The observation or non-observation of the $0\nu\beta\beta$ with high sensitivity could provide crucial insights and impose stringent constraints on the parameter space of the seesaw mechanism.
In our previous study, as shown in Ref.~\cite{Fang:2021jfv}, we have employed the combined effective neutrino mass $m_{\rm eff}$
of the $0\nu\beta\beta$ in the seesaw model by using the empirical mass-dependent relation of the NMEs, and established the bounds on the parameter space given a particular upper limit of $m_{\rm eff}$.
In the current study we shall extend to the constraints on the minimal Type I seesaw model~\cite{Frampton:2002qc,Xing:2020ald,Guo:2006qa} from the current and future $0\nu\beta\beta$ experimental results.
We present a calculation of the mass dependent NMEs from the nuclear many body approach of the quasi-particle random
phase approximation (QRPA). Given that a consensus on the convergence of various many-body methods has yet to be achieved, we also compile the current available calculation results of NMEs from light and heavy Majorana neutrino exchanges, and construct the mass-dependent NMEs from the Interacting Boson Model (IBM), Covariant Density Functional Theory (CDFT) and Interacting Shell Model (ISM) using the empirical interpolation relation~\cite{Faessler:2014kka}. By using the model spread from different many body approaches and considering the correlation among the NMEs of different Majorana neutrino masses, we derive the constraints on the parameter space of the minimal Type-I seesaw model from the current $0\nu\beta\beta$ data and compare with the limits from other experimental probes. We also present the sensitivities on the parameter space from the next generation experiments.

The current work is arranged as follows. In Sec.~\ref{sec:theory}, we introduce the theoretical frameworks for both the nuclear and particle sides, and then an analysis of the mass-dependent NME calculations. In Sec.~\ref{sec:numerical}, we focus on the constraints on the minimal type-I seesaw model parameters from current experiments and the corresponding sensitive ranges from future $0\nu\beta\beta$ experiments. The last section consists of the conclusions and outlook.

\section{Theoretical Framework}\label{sec:theory}
	
 In this section we present the general framework of the $0\nu\beta\beta$ in the type-I seesaw model with two RHNs (i.e., the minimal type-I seesaw model). We first introduce the seesaw relation and the effective neutrino mass, and then present the mass-dependent nuclear matrix elements (NMEs) from the QRPA model. Meanwhile, we also compile the NMEs of exchanging the light and heavy Majorana neutrinos from the literature, and construct the mass-dependent NMEs from other nuclear many body models, such as IBM, CDFT, and ISM, using an empirical interpolation relation~\cite{Faessler:2014kka,Bolton:2019pcu}. 
	
\subsection{The Seesaw relations and the effective neutrino mass}

In the current work, we consider the relevant Lagrangian for the minimal type-I seesaw model as given by~\cite{King:1999mb,Frampton:2002qc,Guo:2006qa,Xing:2020ald}
	\begin{equation}\label{key}
		\mathcal{L} = \mathcal{L}_\mathrm{SM} 
		- \left( Y_{\alpha} \overline{L_\alpha} \Phi N_{\mathrm{R}} + \frac{1}{2} \overline{N_{\mathrm{R}}^c}M_{R}N_{\mathrm{R}} + \rm{h.c.} \right),
	\end{equation}
	where $L_\alpha=(\nu_{\alpha {\rm L}}^{}, l_{\alpha {\rm L}})^\mathrm{T}(\alpha=e,\mu,\tau)$ are the left-handed lepton doublets, $N_{\mathrm{R}}$ denotes the two SM singlets of RHNs and $\Phi$ is the Higgs doublet. $Y_{\alpha}$ is the $3\times2$ neutrino Yukawa coupling matrix and $M_{\mathrm{R}}$ is the $2\times2$ Majorana mass matrix of RHNs. When the scale of the Dirac masses $M_{\mathrm{D}}=Y_{\alpha}\langle \Phi \rangle$ is much smaller than that of the masses of RHNs, the Majorana masses of three light mass eigenstates will be strongly suppressed compared to the scale of the Dirac masses, which is usually called the Seesaw mechanism~\cite{Minkowski:1977sc,Yanagida:1979as,Gell-Mann:1979vob,Mohapatra:1979ia}.
	After spontaneous symmetry breaking, the resultant neutrino mass term is given by
    \begin{equation}
        \mathcal{L}_\mathrm{mass}=
        -\frac{1}{2} 
        \overline{\left( \nu_\mathrm{L}, N_\mathrm{R}^{c} \right)} 
        \left(
        \begin{array}{cc}
             0 & M_\mathrm{D} \\
             M_\mathrm{D}^{\mathrm{T}} & M_\mathrm{R}
        \end{array}
        \right)
        \left(
        \begin{array}{c}
             \nu_\mathrm{L}^{c}  \\
             N_\mathrm{R} 
        \end{array}
        \right)+ \rm{h.c.}\,.
    \end{equation}
The complete $5\times5$ Majorana mass matrix can be decomposed with the unitary mixing matrix ${}\cal U$,
	\begin{equation}
        \left(
        \begin{array}{cc}
             0 & M_\mathrm{D} \\
             M_\mathrm{D}^{\mathrm{T}} & M_\mathrm{R}
        \end{array}\right)
        =
        \left(
        \begin{array}{cc}
             U & R \\
             S & V
        \end{array}\right)
        \left(
        \begin{array}{cc}
             \hat{M}_{\nu} &0 \\
             0 & \hat{M}_{\mathrm{R}}
        \end{array}\right)
        \left(
        \begin{array}{cc}
             U & R \\
             S & V
        \end{array}\right)^{\mathrm{T}}\equiv\mathcal{U}\cdot\hat{\mathcal{M}}\cdot\mathcal{U}^{\mathrm{T}}
        ,
        \label{eq:mass-matrix}
    \end{equation}
	where $\hat{M}_{\nu}=\mathrm{Diag}\{m_1,m_2,m_3\}$ is the diagonalized mass matrix of the three active neutrinos, and $\hat{M}_{\mathrm{R}}=\mathrm{Diag}\{M_1,M_2\}$ includes the two masses of RHNs. $U$, $R$, $S$, $V$ are the upper-left $3\times3$, upper-right $3\times2$, lower-left $2\times3$, lower-right $2\times2$ sub-matrices of the unitary matrix, respectively~\cite{Xing:2007zj,Xing:2011ur}, in which $U$ is the mixing matrix of active neutrinos in the SM charged current (CC) interactions, usually called the PMNS matrix~\cite{Maki:1962mu,Pontecorvo:1957qd}, and $R$ is the mixing matrix of RHNs in the CC interactions. In the type-I seesaw model, $U$ is not unitary because of the non-vanishing $R$, which results in the intrinsic relation between the mass and mixing elements of the seesaw mechanism:
	\begin{equation}\label{eq:intrinsic relation}
		\sum_{i} U_{\alpha i}^2 m_i +  \sum_{I} R_{\alpha I}^2 M_I = 0\;,
	\end{equation}
 where $i= 1,2,3$ and $I=1,2$.
	 The left-handed active flavor neutrinos in the CC interactions are written as the mixing of all the mass eigenstates:
	\begin{equation}\label{mix6}
		\nu_{\alpha \mathrm{L} }=\sum_{i} U_{\alpha i} \nu_{i}+\sum_{I} R_{\alpha I} N_{I}^{c},
	\end{equation}
	where $N_I$ ($I=1,2$) are the mass eigenstates of RHNs.

For the $0\nu\beta\beta$-decay process with the neutrino mass mechanism, only the Standard Model charged-current (CC) interaction is relevant and the effective Hamiltonian is written as
\begin{eqnarray}
\mathcal{H}_{\rm int} = J_{L\mu} j^{\mu}_L\,,
\end{eqnarray}
where $J_{L\mu}$ and $j_{L}^{\mu}$ are the nuclear and lepton weak currents respectively, with the lepton current $j_{L\mu}$ given by
\begin{eqnarray}
j_{L\mu}=\overline{L_\alpha} \tau^+ \gamma_\mu L_\alpha\,,
\end{eqnarray}
where $\tau$ is the isospin operator connecting the charged lepton and neutrino states in the lepton doublet.
For the $0\nu\beta\beta$ process,
the Majorana neutrino mass term enables the contraction of two identical neutrino mass eigenstates, which violates the lepton number by two units. Then the effective Hamiltonian can be divided into three parts:
\begin{eqnarray}
H_{\rm int}=\int d\vec{x} \int d\vec{y} J_{L\mu}(\vec{x}) J_{L\nu}(\vec{y}) \bar{\psi}_{eL}(\vec{x})\gamma^\mu D_{\nu_e} (\vec{x}-\vec{y}) \gamma^\nu \psi_{eL}^C(\vec{y})  \;, 
\end{eqnarray}
the nuclear currents, the neutrino propagator and the electron currents, with $\psi_{eL}^C=C \bar{\psi}_{eL}$ and $C$ being the charge conjugate. Here the neutrino propagator $D_{\nu_e }(\vec{x},\vec{y})$ is superposition of propagators for different mass eigenstates:
\begin{eqnarray}
D_{\nu_e }(\vec{x},\vec{y}) &=&\sum_{i} U_{ei}^2 \int  \frac{d^4 q}{(2\pi)^4} e^{iq(x-y)} \frac{\not \! q+m_i}{q^2+m_i^2} 
\end{eqnarray}
where for weak interactions, both vertices are left-handed, and only the mass term in the propagator is relevant.

The nuclear weak currents can be written under the Breit frame as \cite{Simkovic:1999re}:
\begin{eqnarray}
J_{\mu}(\vec{x}) &=&\sum_{n=1}^{A} \left[\delta_{0\mu} g_V(q^2) + \delta_{j\mu} \left(g_A(q^2)\sigma_j + \frac{g_M(q^2)}{2m_p} i(\sigma \times \vec{q})_j \right.\right.\nonumber\\ 
&& \left.\left. - \frac{g_P(q^2)}{2m_p}\vec{\sigma}\cdot\vec{q} q_j\right)\right]\delta(\vec{x}-\vec{r}_n)
\end{eqnarray}

The inverse of $0\nu\beta\beta$ half life $T^{0\nu}_{1/2}$ can be obtained from the $S$-matrix theory \cite{Doi:1985dx}:
\begin{eqnarray}
1/T^{0\nu}_{1/2} = G^{0\nu} |{\cal M}^{0\nu}|^2 \;,
\label{half-life1}
\end{eqnarray}
 where the phase space factor $G^{0\nu}$ is the integration over the electron momenta and ${\cal M}^{0\nu}$ denotes the contribution from both the nuclear and lepton interactions. On deriving this expression, several assumptions such as the no-finite de Broglie wave length approximations~\cite{Doi:1985dx} are used to separate the lepton and nuclear parts. In this way, the phase space factors can be obtained numerically with high precision~\cite{Kotila:2012zza}.

\subsection{Mass-dependent NME calculations}

    The $0\nu\beta\beta$ NME in the minimal type-I see-saw model can be divided into different parts of the Fermi (F), Gamow-Teller (GT), and tensor transitions according to the different components of the weak currents.  In 2018, a new important short-range interaction as the leading contribution is identified due to renormalization arguments ~\cite{Cirigliano:2018yza, Cirigliano:2018hja,Wirth:2021pij,Jokiniemi:2021qqv,Cirigliano:2020dmx}. 
    Recently, 
    Ref. \cite{Liu:2024kwc} suggests that this term is numerically equivalent to the contribution from form factor which we already considered in this work,
    therefore we temporarily neglect its contribution in current discussion. 
    Meanwhile, the latest chiral EFT study \cite{Dekens:2023iyc,Dekens:2024hlz} identifies an important ultra-soft contribution from light-mass neutrino, which we also neglect for now.
    On the other hand, in current work, we give a general analysis for different traditional many-body approaches, while for most methods, the estimations of this new term are not sufficient. Therefore, before we could properly account for this new contribution in most many-body calculations, we temporarily neglect its contribution in our discussion.
    
    The NME for current calculation can be derived from the combination of the nuclear weak current and the neutrino propagators mentioned above:
    \begin{eqnarray}
    {\cal M}^{0\nu}=\sum_{IJ,\alpha,jk} \mathcal{U}_{ej} \mathcal{U}_{ek} M_{jk}^{IJ,\alpha}\,,
    \end{eqnarray}
    where the capital letters $I,J$ refer to the different components of weak currents (Axial current(A), weak magnetism (M), {\it etc.}) and small letters $j,k$ refer to the neutrino mass eigenstates, and $\alpha$ are different combinations of the nuclear current, which can be divided into 3 parts: $F$ from the time-time components of the nuclear currents ($\mathcal{O}_F=1$) and $GT$ ($\mathcal{O}_{GT}=\sigma_1 \cdot \sigma_{2}$) as well as $T$ ($\mathcal{O}_T=3(\vec{\sigma}_1\cdot\vec{q})(\vec{\sigma}_2\cdot\vec{q})-\vec{\sigma}_1 \cdot \vec{\sigma}_2$) from the space-space components of the currents. Within the minimal type-I see-saw mechanism, the matrix elements can be written as:
    \begin{eqnarray}
    M_{IJ,\alpha,jk}= \sum_{i,f} \delta_{jk} m_j \langle f|| \mathcal{O}_{\alpha} H_{IJ,\alpha}^{j} ||i\rangle \;.
    \end{eqnarray}
    Here $H$ is the so-called neutrino potential which is actually the neutrino propagator:
    \begin{eqnarray}
    H_{IJ,\alpha}^{j}=\int \frac{d\vec{p}}{(2\pi)^3} \frac{e^{i\vec{q}\cdot(\vec{x}-\vec{y})}}{\omega_i(\omega_i+A_N)} h_{IJ,\alpha}(q^2)\;,
    \end{eqnarray}
    where $\omega_{j}=\sqrt{q^2+m_j^2}$ with $m_j$ stands for arbitrary neutrino mass. Here the first term in the energy denominator comes from the contour integration over $q^0$ of the neutrino propagator and the second term from the integration over the time coordinate in deriving the S-matrix. To get the second term, one assumes the electrons share the decay energy, $A_N=E_{xN}+M_m-(M_i+M_j)/2$, where $E_{xN}$ is the excitation energy of the Nth intermediate state, $M_i$, $M_j$ and $M_m$ standing for the nucleus masses from the initial, final and intermediate state, respectively. 
    
    The above NME is neutrino mass dependent. As one may be aware, the typical exchange momentum of neutrino is around several tens to hundreds of MeV in nuclear environment, generally the magnitude of $\pi$ mass, denoted as $\langle q \rangle$. If neutrino mass differs largely with the neutrino exchange momentum, it can extracted out and the NME becomes mass independent. For example, if $m_j \ll \langle q \rangle$, then $\omega_j \approx q$, then $M_{IJ,\alpha,jk}\sim \delta_{jk} m_j M_{IJ,\alpha,{\rm L}}$ while if $m_j \gg \langle q \rangle$, we have $M_{IJ,\alpha,jk}\sim \delta_{jk} M_{IJ,\alpha,{\rm H}}/m_j$, here L and H refer to light and heavy respectively (In literature, the heavy neutrino case is actually defined as $M_{IJ,\alpha,jk}\equiv \delta_{jk} m_p m_e M_{IJ,\alpha,{\rm H}}/m_j$ for the dimension issue, and we follow this definition in our subsequent discussions). These two cases actually correspond to the light and heavy mass mechanism \cite{Simkovic:1999re}. The mass dependent NME for the whole neutrino mass range can in principle be calculated microscopically with any nuclear many-body approaches and has been done before for spherical Quasi-particle Random Phase Approximation (sQRPA) calculations \cite{Faessler:2014kka}. Since most methods involved in our analysis have not provided the mass dependent NME, one usually resorts to the interpolation formula \cite{Faessler:2014kka} \footnote{Note that a new functional form approximately describing the mass dependence of NME for $^{76}{\rm Ge}$ and $^{136}{\rm Xe}$ has been proposed recently in Refs. \cite{Dekens:2023iyc,Dekens:2024hlz}, which aims at higher accuracy and includes a linear term of $m_j$ under the chiral effective field theory (EFT) framework \cite{Cirigliano:2018yza}.}:
    \begin{eqnarray}
    M_{0\nu}(m_j)=\frac{m_p m_e}{\langle p^2 \rangle+m_j^2} M_{\rm H}\;,
    \label{eq:NME-interpolation}
    \end{eqnarray}
    where $\langle p^2\rangle =m_p m_e M_{\rm H}/M_{\rm L}$ with the total NME in the light neutrino exchange mechanism as $M_{\rm L} = \sum_{IJ,\alpha}M_{IJ,\alpha,L}$ and 
    the total NME in the heavy neutrino exchange mechanism as $M_{\rm H} = \sum_{IJ,\alpha}M_{IJ,\alpha,H}$.
    
    For the calculations of NME, it can actually be divided into two parts for most many-body approaches, the single particle matrix elements and the nuclear transition density:
    \begin{eqnarray}
    M_{IJ,\alpha,jk} & = & \sum_{p_1 p_2 n_1 n_2 }^{mK} \langle p_1 n_1 K|| \mathcal{O}_{IJ,\alpha,jk} || p_2 n_2 K\rangle \langle 0_f^+ || [c_{p_1}^\dagger \tilde{c}_{n_1}]_K|| m\rangle \langle m || [c_{p_2}^\dagger \tilde{c}_{n_2}]_K|| 0_i^\dagger \rangle   \nonumber\\ 
    &=& \sum_{p_1 p_2 n_1 n_2 }^{\mathcal{J}} \langle p_1 p_2 \mathcal{J} || \mathcal{O}_{IK,\alpha,jk} || n_1 n_2 \mathcal{J} \rangle \langle 0_f^+ || [c_{p_1} c_{p_2}]_{\mathcal{J}} \cdot [\tilde{c}_{n_2} \tilde{c}_{n_1}]_{\mathcal{J}} ||0_i^+\rangle\;,
    \label{nme}
    \end{eqnarray}
    where K denotes the spin-parity of the intermediate states $| m\rangle$ , and $\cal{J}$ represents the total angular momentum of decaying nucleon pair.
    The second line applies to the approaches using the closure approximation which doesn't explicitly include the intermediate states. This is used for most approaches whose results are involved in current work. While the first line is used for the method which calculates explicitly the intermediate state, mostly the QRPA method. The single particle matrix elements (the first terms in the rhs of above equation) for the two cases can be transformed into each other with the angular momentum algebra~\cite{Simkovic:2007vu}.

For simplicity, we rewrite the inverse half life in Eq. (\ref{half-life1}) as \cite{Fang:2021jfv}
\begin{equation}
    1/T^{0\nu}_{1/2} = G |M_{0\nu} (0) \cdot m_{\rm eff}|^2\;.
    \label{eq:half-life}
\end{equation}
Here the redefined phase factor $G$ is in unit of ${\rm eV}^{-2} {\rm yr}^{-1}$ by dividing $m_e^2$ and absorbing $g_{A0}^4$ ($g_{A0}$ denotes the bare value of axial vector coupling constant) from ${\cal M}^{0\nu}$,
and 
\begin{equation}
    |m_{\rm eff}| = \left||m_{\rm eff}^{\nu}| - |m_{\rm eff}^{\nu}|f_{\beta} (M_2) + |R_{e1}^2| e^{2{\rm i} \delta_{14}} M_1 \left[f_{\beta} (M_1) - f_{\beta} (M_2)\right] \right|\;.
    \label{eq:meff}
\end{equation}
where $R_{e1}$ is the $(1,1)$ element of matrix $R$ in Eq. (\ref{eq:mass-matrix}), $2\delta_{14}$ denotes the relative phase of $R_{e1}^2$ and $m_{\rm eff}^{\nu}$ ranging from $0$
to $2 \pi$, and $f_{\beta}(M_N) = M_{0\nu} (M_N) / M_{0\nu} (0)$ with $M_{0\nu} (M_N)$ being the overall NME dependent of arbitrary neutrino mass $M_N$ in consideration.
Moreover, the effective active neutrino mass $m_{\rm eff}^{\nu}$ is expressed as
\begin{eqnarray}
    |m_{\rm eff}^{\nu}| &=&\left|\sum_{i=1}^3 m_i U_{ei}^2 \right| \nonumber\\
    &=& \left|m_1 \cos^2\theta_{13} \cos^2 \theta_{12} + m_2 \cos^2 \theta_{13} \sin^2 \theta_{12} e^{{\rm i} \alpha_{2}} + m_3 \sin^2 \theta_{13} e^{{\rm i} \alpha_3}\right|
\end{eqnarray}
by considering the standard parameterizaion of $U$,
where $m_i$ (for $i=1,2,3$) are the masses of three kinds of active neutrinos, $\theta_{12}$ and
$\theta_{13}$ the solar and reactor neutrino mixing angles, respectively, $\alpha_2$ and $\alpha_3$ two Majorana phases. In the minimal type-I seesaw with the lightest neutrino mass being zero, we have $m_1=0$ for the normal mass ordering (NMO, $m_1<m_2<m_3$) of neutrino masses, and $m_3=0$ for the inverted mass ordering (IMO, $m_3<m_1<m_2$) of neutrino masses. This directly leads to $|m_{\rm eff}^{\nu}| =|\sqrt{\Delta m^2_{21}} \cos^2 \theta_{13} \sin^2 \theta_{12} + \sqrt{\Delta m^2_{31}}\sin^2 \theta_{13} e^{{\rm i} (\alpha_3-\alpha_2)}|$ for the NMO and $|m_{\rm eff}^{\nu}|
    =|\sqrt{|\Delta m^2_{31}|} \cos^2\theta_{13} \cos^2 \theta_{12} + \sqrt{|\Delta m^2_{32}|} \cos^2 \theta_{13} \sin^2 \theta_{12} e^{{\rm i} \alpha_{2}} |$ for the IMO, where $\Delta m_{ij}^2 = m_i^2 - m_j^2$ for $ij=21,31,32$. About $m_{\rm eff}^{\nu}$, the largest uncertainty comes from the Majorana phases and other parameters have been measured with a good precision by current neutrino oscillation experiments \cite{Esteban:2020cvm}.

\subsection{NME calculations from various nuclear many-body approaches}
    
\begin{figure}
		\centering\includegraphics[width=0.8\textwidth]{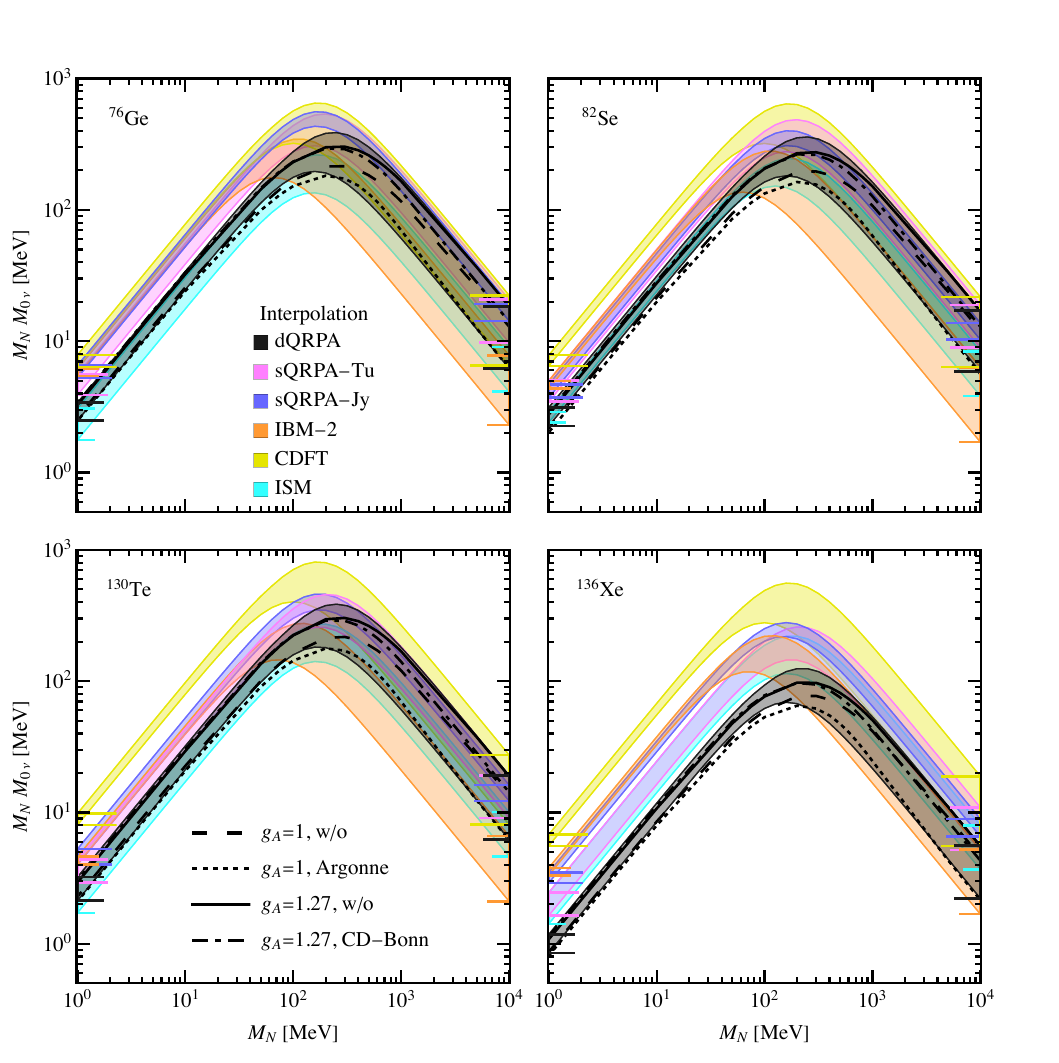}
		\caption{The change of NME value with different neutrino masses in the case of different NME models, where the filled bands are inferred based on the empirical interpolation formula in Eq. \eqref{eq:NME-interpolation} while the black lines are derived from detailed simulations.}
		\label{fig:NME}
	\end{figure}

In the current work, we use the deformed QRPA (dQRPA) method with realistic forces \cite{Fang:2018tui} for the explicit calculations of mass dependent NMEs. We will not go into much detail of the calculation. The major parameters of this method are the particle-particle interaction strength $g_{pp}$'s, which are usually fixed by reproducing the experimental NMEs \cite{Simkovic:2013qiy}. The uncertainty also arises from the quenching of axial vector coupling constant $g_A$, as observed in charge exchange experiments. In our work, we use two sets of $g_A$ values: the bare $g_{A0}=1.27$ obtained from neutron decay \cite{ParticleDataGroup:2020ssz} and $g_A=0.75g_{A0}\simeq1$ by comparing the measured Gamow-Teller strength with corresponding sum rules 
\cite{Towner:1987zz}.
These two sets of results usually set the upper and lower boundaries of the $0\nu\beta\beta$ NME. Other factors that affect the NME calculation are the residual interactions and the related short-range correlation (src). It is found in Refs.~\cite{Simkovic:2013qiy,Fang:2018tui} that NMEs are barely affected by the residual interactions. But src can be important if the neutrino exchange momentum is large. Thus the uncertainty from src will lead to the errors of NME calculation especially when the neutrino mass $M_N$ is large, as shown in Fig.~\ref{fig:NME}.

Such calculations have also been done in the spherical cases~\cite{Faessler:2014kka}. In the current work, for spherical QRPA (sQRPA) calculations, we use only the interpolation formula to obtain the neutrino-mass dependent NME rather than explicit simulations. The current approach of dQRPA calculations differs from that of ``sQRPA-Tu" from Ref.~\cite{Faessler:2014kka} and that of ``sQRPA-Jy" from Ref.~\cite{Hyvarinen:2015bda} by the ground state wave functions. In our case, we use the deformed Woods-Saxon meanfield while the latter ones use a spherical meanfield. And as pointed out in Ref.~\cite{Fang:2011da}, deformed calculation gives suppression to the NME mostly from the overlap factors between the initial and final meanfields which are ignored in the spherical calculations. Therefore, as we can see from Tables~\ref{tab:LNE} and \ref{tab:HNE}, at the small and large neutrino mass ends, we have always smaller NMEs than the spherical calculations.
    
The NMEs can also be calculated by several other many-body approaches, i.e., IBM-2~\cite{Barea:2013bz}, CDFT~\cite{Song:2017ktj} and ISM~\cite{Menendez:2017fdf}, of which IBM-2 differs from the others in the means that the effective nuclear force is not explicitly incorporated. In the current work, the {\it ab inito} approaches for which the nuclear forces obtained from bare nucleon-nucleon interaction \cite{Hergert:2015awm, Novario:2020dmr, Yao:2020olm,Hergert:2020bxy,Ekstrom:2022yea,Gomez-Cadenas:2023vca} are not included. We use another category of the nuclear force, called phenomenological force whose parameters are fixed by reproducing the nuclear bulk properties. 

Unlike the QRPA approach, most methods mentioned here adopt the so-called closure approximation which treats the complete set of intermediate states as the unity matrix. With such an approximation, only the initial and final ground states are needed for the calculation. In this case, what we need for the nuclear input is hence the two body transition density (TBTD) between the two ground states, and a closure intermediate state energy which is the average excitation energy for all the intermediate states. The general expression for the NME therefore follows the second line of Eq. \eqref{nme}. Thus the difference of these methods stems from the ground states wave functions which are solutions of the many-body approaches. 

The Interacting Shell Model (ISM) takes advantage of the existence of the large shell gaps and divides the particles into two parts: the core and valence particles. The corresponding nuclear interactions are obtained from fitting the nuclear properties, such as excitation energies and transition strength. It is regarded as the exact solution of the nuclear many-body problem providing the nuclear interactions are accurate enough. However, it suffers from the huge requirement of computing resources, and as a compromise, for the heavier nuclei, severe model space truncation is usually used.  

Meanwhile the (Covariant) Density Function Theory (DFT or CDFT) methods \cite{Song:2017ktj} usually base on the self-consistent Hatree-Fock Boglyubov (HFB) meanfield and can deal with the shape coexistence which is missing in QRPA or the Projected Hatree-Fock Boglyubov (pHFB) methods \cite{Rath:2013fma}. This is done by treating the deformation as the collective coordinate and obtaining the ground states energies by the variation method. Besides the deformation, other variables can also be treated as the collective coordinates, such as the iso-scalar pairing \cite{Hinohara:2014lqa} in particle-particle channel. In addition, in this work we take the results from the Interacting Boson Model (IBM-2), which starts from a quite different points compared with the above nuclear many-body approaches.

In the current work, we adopt the results from the above discussed approaches published in various literatures~\cite{Simkovic:2013qiy,Hyvarinen:2015bda,Menendez:2008jp,Barea:2013bz,Song:2017ktj}, and they are presented in Tables~\ref{tab:LNE} and \ref{tab:HNE}. Two sources of external errors which are common issues for different methods are included in the current study: the choices of the value of $g_A$ (two cases: bare and quenched $g_A$ extracted from various experiments) which probably originate from the many-body current \cite{Gysbers:2019uyb} and the srcs (in most literatures, four cases are used: bare, Argonne, CD Bonn and Miller-Spencer). For all these different models, we define a common scenario, that is with CD-Bonn src and bare $g_A$. We will discuss the comparison among different models based on this common scenario later in this work. 

While Table~\ref{tab:LNE} and Table~\ref{tab:HNE} present the results for the two extreme cases, we illustrate the neutrino mass dependent NME results of ${}^{76}\mathrm{Ge}$, ${}^{82}\mathrm{Se}$, ${}^{130}\mathrm{Te}$, ${}^{136}\mathrm{Xe}$ used in our analysis from various approaches in Fig.~\ref{fig:NME}. 
Our calculation with dQRPA gives explicitly the neutrino mass dependent NME, while for other methods, the NMEs for different neutrino mass are interpolated from the NMEs for light- ($M_N\rightarrow 0$) and heavy- ($M_N\rightarrow\infty$) mass mechanisms. In this sense, calculations, such as those from deformed Skyrme QRPA \cite{Mustonen:2013zu}, are not included because they have only results for $M_N\rightarrow 0$. And the errors of adopting such interpolating formula have been estimated in~\cite{Faessler:2014kka} to be about 20\%-25\% at the region of $M_N\sim \langle q \rangle$, while otherwise the errors are small. This has also been checked in this work. In Fig.~\ref{fig:NME}, the explicit calculations (lines) and interpolated NMEs (shades) are given together. While they overlap with each other for tiny or large $m_N$, the interpolated results always over-predict the relative NME with $M_N \lesssim m_\pi$ and under-predict the NME if $M_N\gtrsim m_\pi$. For the case without quenching and src, the uncertainties brought by the interpolation are around 50\%. If the $g_A$ quenching and src are accounted, this error is brought down by around ten percents.   

\begin{table}[]
    \centering
    \resizebox{\columnwidth}{!}{%
    \begin{tabular}{c|cc|cccccc}
    \hline
    \hline
 &  $g_A$ & src & dQRPA \cite{Fang:2018tui} & sQRPA-Tu \cite{Simkovic:2013qiy} & sQRPA-Jy \cite{Hyvarinen:2015bda} & IBM-2 \cite{Barea:2015kwa} & CDFT \cite{Song:2017ktj} & ISM \cite{Menendez:2017fdf}  \\
    \hline
\multirow{7}*{$^{76}$Ge} & 1.27 & w/o  & 3.27  &   -     &   -   &   -   & 7.61  & -\\
 &            & Argonne & 3.12  & 5.157   &    -   & 5.98  & 7.48  & 2.89  \\
 &          & CD-Bonn   & 3.40  & 5.571   & 6.54  & 6.16  & 7.84  & 3.07  \\
 &          & Miller-Spencer        &    -   &      -   &    -   & 5.42  & 6.36  & - \\
        & 1.00 & w/o    & 2.64  &   -       & -   & -    & -     &   -   \\
 &      & Argonne       & 2.48  & 3.886     & -   &   -   &   -   &   1.77   \\
 &      & CD-Bonn       & 2.72  & 4.221     &5.26 & -   &   -   &   1.88   \\
 \hline
 $^{82}$Se & 1.27 & w/o & 3.01  & -     & -     & -      & 7.60 &   -\\
        &  & Argonne    & 2.86  & 4.642 &  -    & 4.84  & 7.48  &   2.73  \\
 &          & CD-Bonn   & 3.13  & 5.018 & 4.69  & 4.99  &7.83  &  2.90  \\
 &          & Miller-Spencer    &   -  & -  & - & 4.37  & 6.48  & -      \\
 & 1.00     & w/o       & 2.41  &  -    & -     & -     & -     & -   \\
 &          & Argonne   & 2.26  & 3.460 & -     & -     & -     & 2.41  \\
 &          & CD-Bonn   & 2.49  & 3.746 & 3.73  & -     & -     & 2.56 \\
  \hline
 $^{130}$Te & 1.27& w/o & 3.10  &       &       &       & 9.55 &  \\
 &          & Argonne   & 2.90  & 3.888 &       & 4.47  & 9.38 & 2.76 \\
 &          & CD-Bonn   & 3.22  & 4.373 & 5.27  & 4.61  & 9.82 & 2.96 \\
 &          & Miller-Spencer    & - & - &  -    & 4.03  & 8.03 & \\
            & 1.00 & w/o& 2.29  &    \\
 &          & Argonne   & 2.13  & 2.945 & -     &  -    &  -    &   1.72  \\
 &          & CD-Bonn   & 2.37  & 3.297 & 4.00  &  -    &   -   &   1.84  \\
  \hline
 $^{136}$Xe & 1.27& w/o & 1.12  &  -    &  -    &  -    &  6.62 &\\
 &          & Argonne   & 1.11  & 2.177 &       & 3.67  &  6.51 & 2.28\\
 &          & CD-Bonn   & 1.18  & 2.460 & 3.50  & 3.79  &  6.80 & 2.45\\
 &          & Miller-Spencer& - &   -   &  -    & 3.33  &  5.58 &\\
            & 1.00& w/o & 0.85  &  & \\
 &          & Argonne   & 0.86  & 1.643 &   -   &   -   &   -   &   1.42\\
 &          & CD-Bonn   & 0.89  & 1.847 & 2.91  &   -   &   -   &   1.53\\
    \hline
    \hline
    \end{tabular} %
    }
    \caption{Nuclear matrix elements for light neutrino exchange mechanism $M_{\rm L}$ obtained by various nuclear many-body models.}
    \label{tab:LNE}
\end{table}

\begin{table}[]
    \centering
    \resizebox{\columnwidth}{!}{%
    \begin{tabular}{c|cc|cccccc}
    \hline
    \hline
 &  $g_A$ & ${\rm src}$ & dQRPA \cite{Fang:2018tui} & sQRPA-Tu \cite{Simkovic:2013qiy} & sQRPA-Jy \cite{Hyvarinen:2015bda} & IBM-2 \cite{Barea:2015kwa} & CDFT \cite{Song:2017ktj} & ISM \cite{Menendez:2017fdf}  \\
    \hline
\multirow{7}*{$^{76}$Ge} 
        & 1.27 & w/o    & 385.4  &      &       &       & 466.8  & \\
 &            & Argonne & 187.3  & 316  &       & 107  & 267  & 130 \\
 &          & CD-Bonn   & 293.7  & 433  & 401.3 & 163  & 378.1  & 188  \\
 &          & Miller-Spencer &   &       &      & 48.1  & 135.7  &  \\
        & 1.00 & w/o    & 275.9  &\\
 &      & Argonne       & 129.7  & 204  &       &       &       &   86\\
 &      & CD-Bonn       & 207.2  & 287  & 298.3 &       &       &   122\\
 \hline
 $^{82}$Se & 1.27 & w/o & 358.7 &       &       &       & 454  &    \\
        &  & Argonne    & 175.9 & 287   &       & 84.4  & 261.4   &   121 \\
 &          & CD-Bonn   & 273.6 & 394   & 287.1 & 132  &  369  &   175\\
 &          & Miller-Spencer        &   &   &   & 35.6  & 132.7 &   \\
 & 1.00     & w/o       & 257.4 &     \\
 &          & Argonne   & 122.1 & 186   &   -   &   -   &   -   &   80  \\
 &          & CD-Bonn   & 193.4 & 262   &  214.3 &   -   &   -   &   113 \\
  \hline
 $^{130}$Te & 1.27& w/o & 401.1 &       &       &       &   573 & \\
 &          & Argonne   & 191.4 & 292   &       & 92  &   339.2 &   146\\
 &          & CD-Bonn   & 303.5 & 400   & 338.3  & 138  & 472.8 &   210\\
 &          & Miller-Spencer    &   &       &   & 44  & 168.5 & \\
            & 1.00 & w/o& 281.2 &    \\
 &          & Argonne   & 130.2 & 189   & -     &   -   &   -   &   97  \\
 &          & CD-Bonn   & 209.5 & 264   &  255.7 &   -   &   -   &   136 \\
  \hline
 $^{136}$Xe & 1.27& w/o & 117.1 &       &       &       & 394.5& \\
 &          & Argonne   & 66.9  & 166   &       & 72.8  & 234.3 &   116 \\
 &          & CD-Bonn   & 90.5  & 228   & 186.3 & 109  & 326.2  &   167 \\
 &          & Miller-Spencer    &  -& - &   -   & 35.1  & 116.3 & \\
            & 1.00& w/o & 82.7 &  & \\
 &          & Argonne   & 46.3 & 108    &   -   &   -   &   -   &   77\\
 &          & CD-Bonn   & 62.8 & 152    & 137.3 &   -   &   -   &   108\\
    \hline
    \hline
    \end{tabular} %
    }
    \caption{Nuclear matrix elements for heavy neutrino exchange mechanism $M_{\rm H}$ obtained by various nuclear many-body models.}
    \label{tab:HNE}
\end{table}

\begin{figure}[htbp]
	\centering
	\subfigure{
		\includegraphics[width=0.6\textwidth]{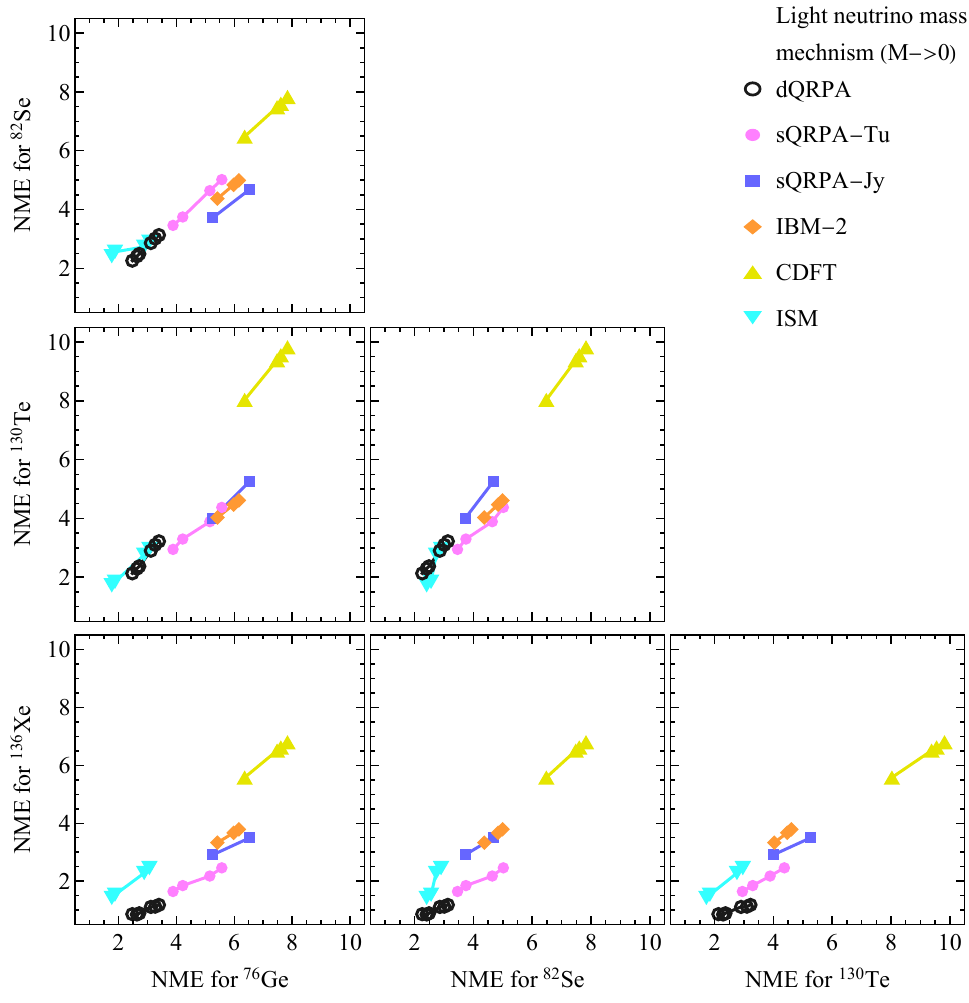}
	}
	\caption{NMEs of several elements in different models for the light neutrino mass mechanism.}
	\label{fig:fig_NME_vs_NME_LNM}
\end{figure}

\begin{figure}[htbp]
	\centering
	\subfigure{
		\includegraphics[width=0.6\textwidth]{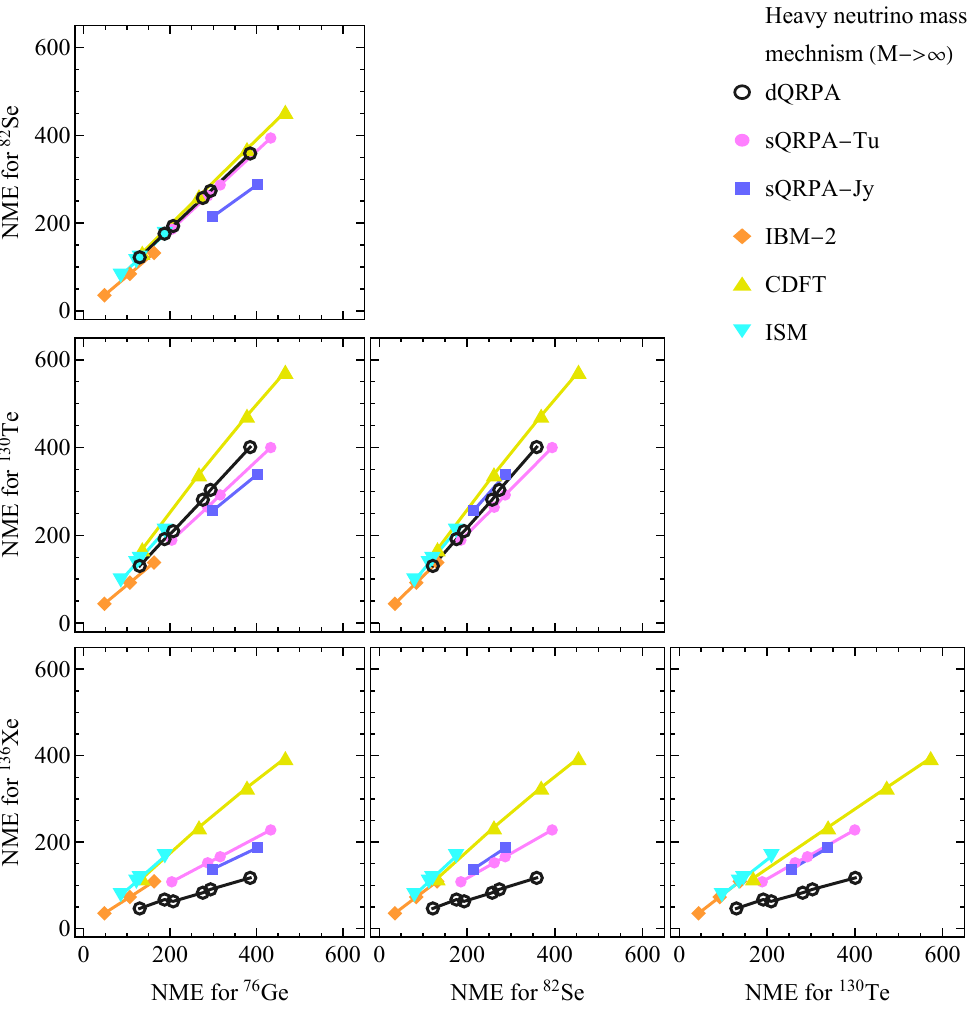}
	}
	\caption{NMEs of several elements in different models for the heavy neutrino mass mechanism.}
	\label{fig:fig_NME_vs_NME_HNM}
\end{figure}

To get a general idea of errors brought by different approaches, we start from a common case mentioned above with CD-Bonn src and bare $g_A$. We find from these tables that for most nuclei, the ISM sets the lower bound of the results for the light mass mechanism while IBM-2 sets the lower bound for the heavy mass mechanism. While for the upper bound, CDFT always has the largest values for the light neutrino mass mechanism, sQRPA gives the upper bound for most nuclei with the heavy neutrino mass mechanism. At the light neutrino mass end, our dQRPA calculation can be even smaller than the ISM results for semi-magic $^{136}$Xe. This may come from the number non-conservation nature of QRPA methods (for sQRPA, the overlap factor between the parent and daughter nuclei is not included, therefore this effect will not appear). It is found that the restoration of particle number conservation will enhance the NMEs for semi-magic nuclei as from the DFT calculations~\cite{Yao:2014uta}. While for light neutrino mass mechanism, model space dependence is not important, the heavy neutrino mass case, the NME is closely related to model space \cite{Brown:2015gsa}. This explains why the dQRPA and ISM give similar results at one end and ISM gives much smaller results at the other end. Alternatively, the major difference between QRPA and CDFT is the inclusion of so-called iso-scalar pairing. This correlation is supposed to be responsible for the suppression of NMEs of both $2\nu\beta\beta$ and $0\nu\beta\beta$ processes \cite{Hinohara:2014lqa}. But one also finds that this sensitivity applies only for light neutrino mass mechanism and the heavy neutrino mass mechanism depends less on this correlation \cite{Fang:2018tui}. This also explains the large discrepancy for CDFT and QRPA at one end but their similarity at another end. Except for $^{136}$Xe, for microscopic models, the general trend is at light neutrino mass mechanism case, ISM with more many-body correlations may be more close to the actual NME but for heavy neutrino mass mechanism, models with large model space may be more reliable. In this sense, dQRPA calculations which agree with ISM for light neutrino mass mechanism and with CDFT for heavy neutrino mass mechanism appear to be reliable, although a careful model benchmark is needed before a concrete conclusion. In general, the discrepancies from various methods can be explained qualitatively \cite{Agostini:2022zub}, but quantitative calculations need to be performed to pin down the deviations as have done in \cite{Brown:2015gsa}.

These analyses are well illustrated by Fig. \ref{fig:fig_NME_vs_NME_LNM} the light neutrino mass mechanism case and Fig. \ref{fig:fig_NME_vs_NME_HNM} the heavy neutrino mass mechanism case as well as in Fig. \ref{fig:NME} for the whole neutrino mass region. Due to the availability of calculated results, the different color bands for different methods are not equal-footing. For all QRPA methods, we have results for both quenched and bare case of $g_A$, but for IBM-2 and CDFT, only results of the bare case $g_A$ are available. As we stated above, the two external sources for uncertainty commonly discussed which is model independent are the values of $g_A$ and the corrections from short range behaviour of nuclear force. A naive estimation of the possible errors from $g_A$ is about certain given that $M_{F}\lesssim 1/3 M_{GT}$ and $M_T$ is small. Under such assumptions, a choice of $g_A$ around 0.75 gives an uncertainty of about 20\% or less. What's more, for light neutrino mass mechanism, we conservatively choose milder src corrections giving a correction of about $5\%\sim10\%$ to the NME, while wilder Miller-Spencer src or even wilder ones are discussed in Refs. \cite{Weiss:2021rig,Sarkar:2023vdl}. 
Therefore for light neutrino case, the quenching may be important for the uncertainty estimation. While for the heavy neutrino mass case, the neutrino potential behaves like a contact potential. Therefore the short-range behaviour of nucleons becomes much more important. For milder CD-Bonn src, a 20\%$\sim$30\% reduction is expected and this reduction could be enhanced to 30\%$\sim$50\% for Argonne src and more than 50\% for Miller-Spencer src. Even without the wildest Miller-Spencer src, the uncertainty brought by src is the most significant source for errors for heavy neutrino mass. In general, the errors from these external sources are quantifiable. That is, the errors are in general 20\% at light neutrino end and a modest estimation with inclusions of more modern src will restrict the errors at heavy neutrino end by 50\%. The errors for arbitrary neutrino mass therefore lie in between 20\% to 50\% if quenching and srcs are considered.
Here we will mainly focus on different NME calculations from various nuclear many body approaches on the market and calculate their phenomenological consequences, respectively. Some NME calculations may be not included due to lacking results for isotopes in consideration. In this way, the NME errors may be bigger.  
However, from the statistical viewpoint,
the actual errors are complicated and related to many aspects of nuclear many-body calculations. One can refer to some efforts in this regard in Refs. \cite{Horoi:2023uah,Horoi:2022ley,Jokiniemi:2022ayc,Belley:2023lec}.

\section{Numerical Results} \label{sec:numerical}
\subsection{The parameter space of effective neutrino mass revisited}
\begin{figure}
    \centering
    \includegraphics[width=0.8\textwidth]{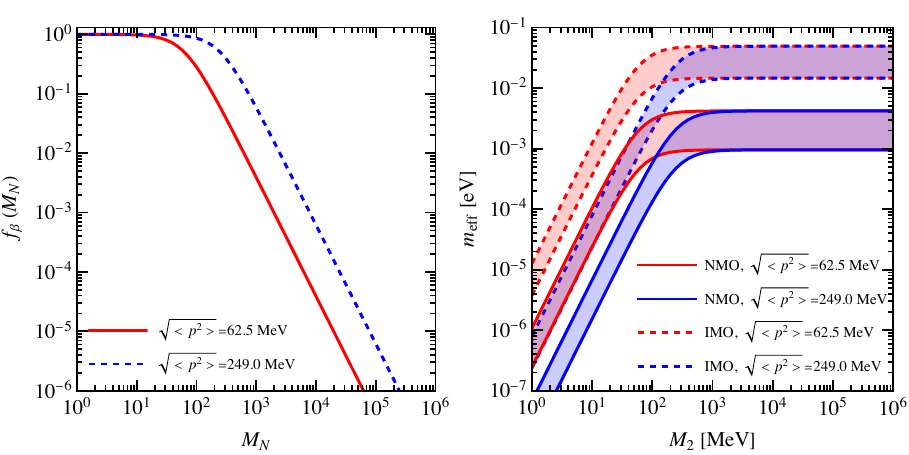}
    \caption{Left panel: the illustration of how $f_{\beta} (M_{N})$ changes with $M_N$; right panel: $m_{\rm eff}$ changing with $M_2$ in the limit $M_1=M_2$. In both figures, the interpolation formula in Eq. \eqref{eq:NME-interpolation} and two specific values of $\sqrt{<p^2>}$ are used.}
    \label{fig:fbeta}
\end{figure}

\begin{figure}
    \centering
    \includegraphics[width=\textwidth]{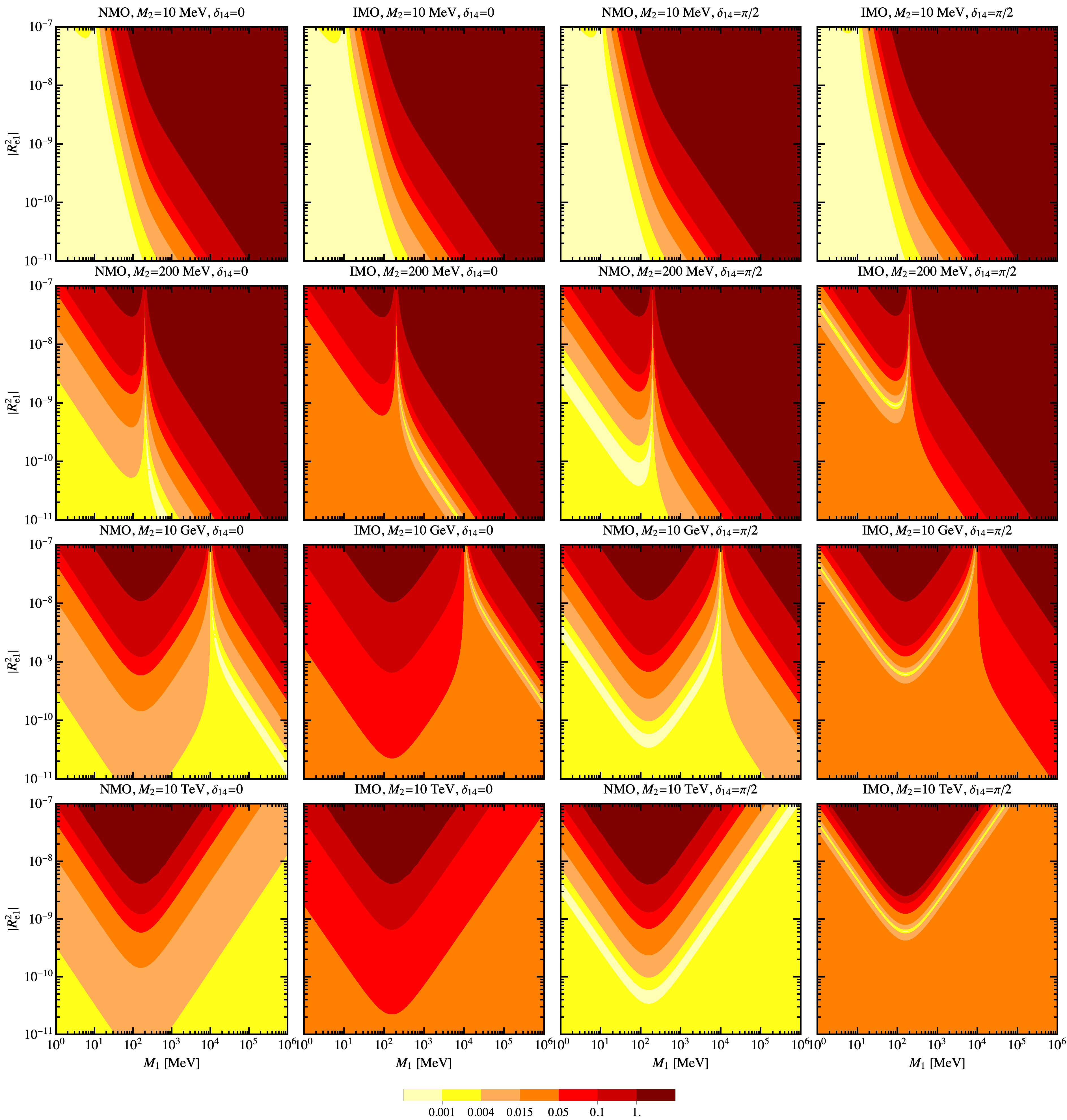}
    \caption{The contour plot of $m_{\rm eff}$ in the dimensions of $M_1$ and $|R_{e1}^2|$ by typically taking
    $\delta_{14}=0$ or $\pi/2$, $\alpha_2-\alpha_3=0$ for NMO, $\alpha_2=0$ for IMO, the NME values in the dQRPA model with $g_A=1$ and Argonne src, as well as the best-fit values of related neutrino oscillation parameters \cite{Esteban:2020cvm}, where the $m_{\rm eff}$ values in the legend are in unit of eV.}
    \label{fig:meff-M1-Re1}
\end{figure}

\begin{figure}
    \centering
    \includegraphics[width=\textwidth]{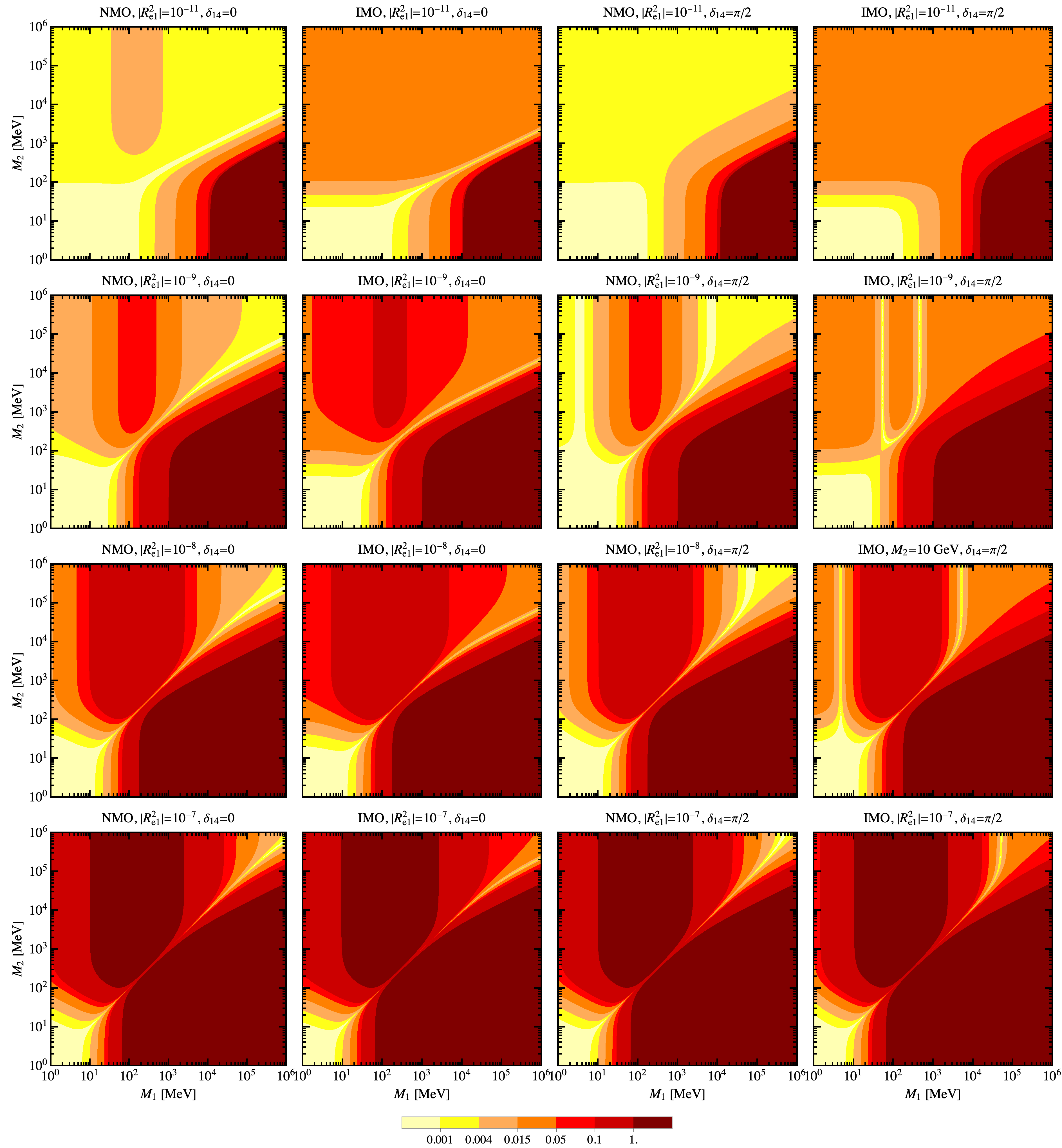}
    \caption{The contour plot of $m_{\rm eff}$ in the dimensions of $M_1$ and $M_2$ by typically taking
    $\delta_{14}=0$ or $\pi/2$, $\alpha_2-\alpha_3=0$ for NMO, $\alpha_2=0$ for IMO, the NME values in the dQRPA model with $g_A=1$ and Argonne src, as well as the best-fit values of related neutrino oscillation parameters \cite{Esteban:2020cvm}, where the $m_{\rm eff}$ values in the legend are in unit of eV.}
    \label{fig:meff-M1-M2}
\end{figure}

As shown in Eq.~\eqref{eq:meff}, the effective neutrino mass $m_{\rm eff}$ depends on five parameters, $m_{\rm eff}^{\nu}$, $M_1$, $M_2$, $|R_{e1}^2|$ and
$\delta_{14}$. In Refs.~\cite{Fang:2021jfv, Dekens:2024hlz}, the authors have studied how the involvement of right-handed neutrinos can magnify or decrease the magnitude of effective neutrino mass. 
We revisit the parameter space of $m_{\rm eff}$ in more details. Considering $3\sigma$ ranges from the current global fit of neutrino oscillation data and allowing the Majorana phases $\alpha_2$ and $\alpha_3$ changing from $0$ to $2 \pi$, we have $0.96~{\rm meV}\leq m_{\rm eff}^{\nu} \leq 4.2~{\rm meV}$ for NMO
and $14.7~{\rm meV}\leq m_{\rm eff}^{\nu} \leq 49.2~{\rm meV}$ for IMO. The whole IMO range will be explored by the next-generation $0\nu\beta\beta$ experiments while the NMO range needs more ambitious proposals beyond.
Having the mixed contribution from the heavy neutrino parameters $(M_1, M_2, |R_{e1}^2|,$ and $ \delta_{14})$, it's difficult to discriminate the neutrino mass orderings from
$0\nu\beta\beta$ experiments. Assuming a fixed value of $m_{\rm eff}^{\nu}$, the $m_{\rm eff}^{}$ can be as small as 0 or exceed the current experimental limit depending on the heavy neutrino parameters.
In Figs. \ref{fig:meff-M1-Re1} and \ref{fig:meff-M1-M2}, we take the NME interpolation formula in Eq. \eqref{eq:NME-interpolation} and illustrate the contour plot of $m_{\rm eff}$ by typically taking $\delta_{14}=0$ or $\pi/2$, $\alpha_2-\alpha_3=0$ for NMO, $\alpha_2=0$ for IMO, the NME values in the dQRPA model with $g_A=1$ and Argonne src, as well as the best-fit values of related neutrino oscillation parameters \cite{Esteban:2020cvm}.
Some comments are as follows:
\begin{itemize}
\item The dependence of $m_{\rm eff}$ on the heavy neutrino parameters in the NMO or IMO case can be very different because the NMO range of $m_{\rm eff}^{\nu}$ is about one order smaller than its IMO range. The yellow region and
orange region correspond to the NMO range and the IMO range, respectively. While the current $0\nu\beta\beta$ experiments already reach part of the IMO range assuming big end of NME value, the next-generation proposals will detect the whole IMO range space and beyond.
Given some upper limit of $m_{\rm eff}$ from experiments, bigger parameter space will be usually excluded in the IMO case.
\item Same as Ref. \cite{Fang:2021jfv}, $\delta_{14} = 0$ or $\pi/2$
are taken as typical examples. According to Eq. \eqref{eq:meff} and as shown in Figs. \ref{fig:meff-M1-Re1} and \ref{fig:meff-M1-M2}, in the case of $M_1>M_2$, $m_{\rm eff}$ with $\delta_{14}=0$ is smaller than its value with $\delta_{14}=\pi/2$ and the conclusion is opposite in the case of $M_1<M_2$.
\item In the limit of $M_1=M_2$, we have $m_{\rm eff} =|m_{\rm eff}^{\nu} [1- f_{\beta} (M_2)] | $ \cite{Fang:2021jfv}. $f_{\beta} (M_2)$ depend on both $M_2$ and $<p^2>= m_p m_e (M_{\rm H}/M_{L})$, where $\sqrt{<p^2>}$ varies from 62.5 MeV to 249.0 MeV from Tables \ref{tab:LNE} and \ref{tab:HNE}. $m_{\rm eff}$ may take any value from 0 to $m_{\rm eff}^{\nu}$ depending on the input of $M_2$. We display how $f_{\beta}(M_N)$ and $m_{\rm eff}$ in the limit $M_1=M_2$  change with heavy neutrino mass in Fig. \ref{fig:fbeta}, where the bands are due to the uncertainties from $m_{\rm eff}^{\nu}$. With $M_1=M_2$ as a boundary, the
$M_{\rm eff}$ depends on $M_1$, $M_2$ and $|R_{e1}^2|$ very differently especially with bigger $|R_{e1}^2|$, which leads to the peak shape in Fig. \ref{fig:meff-M1-Re1} and the asymmetric feature about the diagonal $M_1=M_2$ line.
\item If $m_{\rm eff}$ is big enough, as the $m_{\rm eff}$ increases, the contribution of $m_{\rm eff}^{\nu}$ decreases due to the small ranges of $m_{\rm eff}^{\nu}$ values. In the dark red region, the effects of $m_{\rm eff}^{\nu}$ is negligible. One can refer to \cite{Fang:2021jfv} for more discussion of the contribution of $m_{\rm eff}^{\nu}$. 
\item  From Fig.
\ref{fig:meff-M1-M2}, the contours of big $m_{\rm eff}$ only appears in the part of $M_1>M_2$ when $|R_{e1}^2|$ is small enough while with 
$|R_{e1}^2|$ being big enough, the contours of $m_{\rm eff}$ consist of
two parts, one in the case of $M_1>M_2$, the other $M_1<M_2$.
$m_{\rm eff}$ in the lower left corner and upper right corners is always not very big, especially for the NMO case. For the former,
it is because $f_{\beta} (M_N)$ is not far away from 1, leading to $m_{\rm eff}^{\nu}$ suppressed and meanwhile $M_1$ is not big enough or very different from $M_2$. For the latter, the smallness
of $m_{\rm eff}$ is mainly because the two heavy neutrino masses are not different enough, where the contribution $m_{\rm eff}^{\nu}$ dominates. 
\item With both $M_1$ and $M_2$ being small enough as well as $M_1\neq M_2$, $m_{\rm eff}$ can be very small, which is far beyond the exploration reach of next-generation $0\nu\beta\beta$ experiments. While when $|R_{e1}^2|$ is big enough and $M_1$ is much bigger than $M_2$, $m_{\rm eff}$ is very big and excluded by current experiments (such as the lower part of the dark red region in Fig. \ref{fig:meff-M1-M2}).
\end{itemize}

\subsection{Constraints from current neutrinoless double beta decay experiments}
\label{sec:result}
More specifically, let us interpret the current
$0\nu\beta\beta$ experiment data in the framework of minimal type-I seesaw model. We mainly focus on the upper limit of the effective neutrino mass $m_{\rm eff}$ and the constraints on heavy neutrino parameters by reasonably combining the current available $0\nu \beta\beta$ data. According to Eq. \eqref{eq:half-life} and the definition of $m_{\rm eff}$ in Eq. \eqref{eq:meff}, the upper limit of $m_{\rm eff}$ in the framework of minimal type-I seesaw is equal to the upper limit of $m_{\rm eff}^{\nu}$ in the light-neutrino exchange mechanism. We use the terminology of the $m_{\rm eff}$ upper limit below.
Currently, the strongest constraints of $0\nu\beta\beta$ are obtained from GERDA~\cite{GERDA:2020xhi}, MAJORANA~\cite{Majorana:2022udl} CUORE~\cite{CUORE:2021mvw}, KamLAND-Zen~\cite{KamLAND-Zen:2022tow}, and EXO-200~\cite{EXO-200:2019rkq} experiments.
Most experimental papers provide only these lower limits of $T^{0\nu}_{1/2}$ at 90\% C.L. instead of its $\Delta\chi^2$ function while the $\Delta\chi^2$ values of KamLAND-Zen and CUORE experiments are directly given by the experiment collaborations in Refs.~\cite{CUORE:2021mvw,KamLAND-Zen:2022tow}.
We construct approximately the $\Delta\chi^2$ functions of GERDA, MAJORANA and CUORE similar to Refs.~\cite{Caldwell:2017mqu,Biller:2021bqx,Capozzi:2021fjo,Lisi:2022nka,Pompa:2023jxc,Lisi:2023amm} by parameterizing  $\Delta\chi^2(1/T_{1/2}^{0\nu}) $ as:
\begin{equation}
 \Delta \chi^2_{r} (1/T_{1/2}^{0\nu}) =a_r (1/T_{1/2}^{0\nu})^{2} +
 b_r (1/T_{1/2}^{0\nu}) + c_r\;,
 \label{eq:chi2}
\end{equation}
where $T^{0\nu}_{1/2}$ is in unit of $10^{25}~\mathrm{yr}$ and the subscript $r$ ($r=$ GERDA, MAJORANA, CUORE, KamLAND-Zen and EX0-200) label different experiments. Note that for KamLAND-Zen experiment, with $\Delta\chi^2$ values only for small $T_{1/2}^{0\nu}$ \cite{KamLAND-Zen:2022tow}, we fit them according to Eq. (\ref{eq:chi2}) to get
$\Delta \chi^2_{r} (1/T_{1/2}^{0\nu})$.
As for the CUORE experiment, we fit the available $\Delta \chi^2_{r} (1/T_{1/2}^{0\nu})$ curve given by \cite{CUORE:2021mvw} in accordance with Eq. (\ref{eq:chi2}) for the convenience of later calculations.

\begin{figure}[!htbp]
	\centering
		\includegraphics[width=0.6\textwidth]{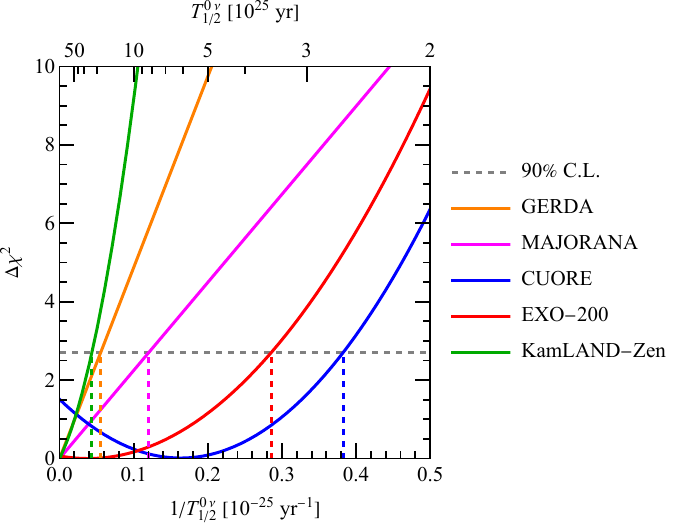}
	\caption{$\Delta\chi^2$ for the various experiments as function of $1/T^{0\nu}_{1/2}$ and $T^{0\nu}_{1/2}$.}
	\label{fig:chi2}
\end{figure}

Table~\ref{tab:Experiments}  shows the derived values of $a_r$, $b_r$ and $c_r$ as well as the corresponding 90\% C.L. ($\Delta\chi^2=2.706$) half-life limits compared with the results published in the experiment papers \footnote{For the CUORE experiment, we take the frequentest limit $T^{0\nu}_{1/2}>2.6\times10^{25}$ yr ($90\%$ C.L.) instead of the Bayesian fit in \cite{CUORE:2021mvw} to use the $\Delta \chi^2$ values there.}. We can see the limits obtained from the $\Delta\chi^2$ function following Eq. (\ref{eq:chi2}) agree well with the published results.
The $\Delta\chi^2 (1/T^{0\nu}_{1/2})$ functions are displayed in Fig.~\ref{fig:chi2}, where we know the most stringent upper limit of $0\nu\beta\beta$ half life is of order $10^{25}$ or $10^{26}$ year at 90\% C.L. dependent of the relevant isotope.

\begin{table}[htb]
    \centering
    \resizebox{\columnwidth}{!}{%
    \begin{tabular}{l|r|r|r|r|c|c|c}
        \hline
        \hline
        Experiment  & Isotope &$a_r$& $b_r$& $c_r$ & $T^{0\nu}_{1/2}/10^{25}$yr (90\%C.L.) & expt. (90\%C.L.) & Reference\\
        \hline
        GERDA       & $^{76}\mathrm{Ge}$ & $0.000$ & 48.708 & 0.000
        &18.0 & 18  & \cite{GERDA:2020xhi}\\
        MAJORANA    & $^{76}\mathrm{Ge}$ & $0.000$ & 22.460 &0.000
        &8.30 & 8.3 & \cite{Majorana:2022udl}\\
        CUORE       & $^{130}\mathrm{Te}$ & $56.041$ & $-18.325$ &1.511
        & 2.61 & 2.6 & \cite{CUORE:2021mvw}\\
        KamLAND-Zen & $^{136}\mathrm{Xe}$ & $527.773$ & 38.780 &0.000
        &22.86 & 23  & \cite{KamLAND-Zen:2022tow}\\
        EXO-200     & $^{136}\mathrm{Xe}$  & $44.300$ & $-3.417$ & 0.066
        &3.50 & 3.5 & \cite{EXO-200:2019rkq}\\
        \hline
        \hline
    \end{tabular}%
    }
    \caption{The coefficients of approximate $\Delta\chi^2 (1/T^{0\nu}_{1/2})$ functions as well as the corresponding 90\% C.L. half-life limits, compared with the publication by the experiment collaborations.}
    \label{tab:Experiments}
\end{table}

In order to fully interpret the existing $0\nu\beta\beta$ results, even though we can not combine all the experiments in
Table \ref{tab:Experiments} to give a valid half life
because the isotope of each experiment is not always the same, we can combine these experimental results to constrain the common part such as the effective neutrino mass $m_{\rm eff}$ and other new-physics parameters involved in $m_{\rm eff}$.
Since these experiments are independent observations,
the combined $\Delta\chi^2$ function for $(T^{0\nu}_{1/2})^{-1}$ is parameterized as summing individual $\Delta\chi^2_r$ function of all $0\nu\beta\beta$ experiments,
\begin{equation}
    \begin{aligned}
    \chi^2_{0\nu\beta\beta} =& 
    \Delta \chi^2_\mathrm{GERDA}[ (T^{0\nu,^{76}\mathrm{Ge}}_{1/2})^{-1} ] +
   \Delta \chi^2_{\mathrm{MAJORANA}}[ (T^{0\nu,^{76}\mathrm{Ge}}_{1/2})^{-1} ] +
    \Delta\chi^2_\mathrm{CUORE}[ (T^{0\nu,^{130}\mathrm{Te}}_{1/2})^{-1} ] \\
    &+
   \Delta \chi^2_{\mathrm{KamLAND-Zen}}[ (T^{0\nu,^{136}\mathrm{Xe}}_{1/2})^{-1} ] +
   \Delta \chi^2_{\mathrm{EXO-200}}[ (T^{0\nu,^{136}\mathrm{Xe}}_{1/2})^{-1} ] 
    \;,
    \end{aligned}
\end{equation}
where $T^{0\nu,^{130}\mathrm{Te}}_{1/2}$, $T^{0\nu,^{136}\mathrm{Xe}}_{1/2}$ and $T^{0\nu,^{76}\mathrm{Ge}}_{1/2}$ are the half-life of the $^{130}\mathrm{Te}$, $^{136}\mathrm{Xe}$ and $^{76}\mathrm{Ge}$, respectively, and can be determined by inputting NME, phase space factor and $m_\mathrm{eff}$ as in Eq. (\ref{eq:half-life}).

\begin{figure}[!htbp]
	\centering
		\includegraphics[width=0.6\textwidth]{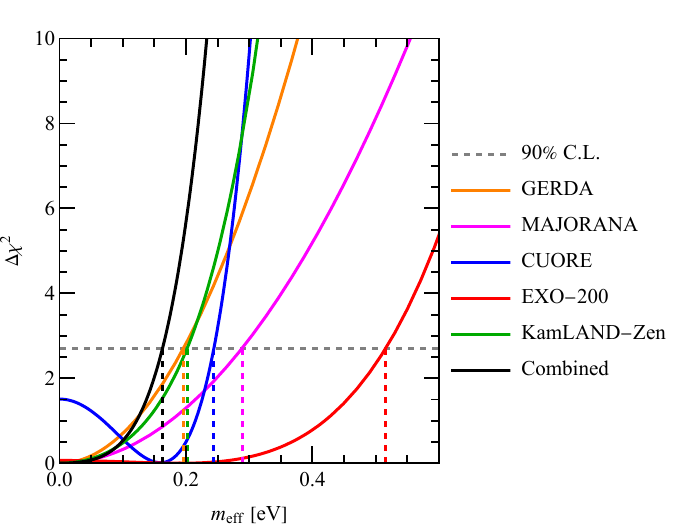}
	\caption{$\Delta\chi^2$ of the $m_\mathrm{eff}$ in dQRPA model ($g_A = 1$ and Argonne src) for the various experiments.}
	\label{fig:chi-meff}
\end{figure}

\begin{figure}[!htbp]
	\centering
	\subfigure{
		\includegraphics[width=0.6\textwidth]{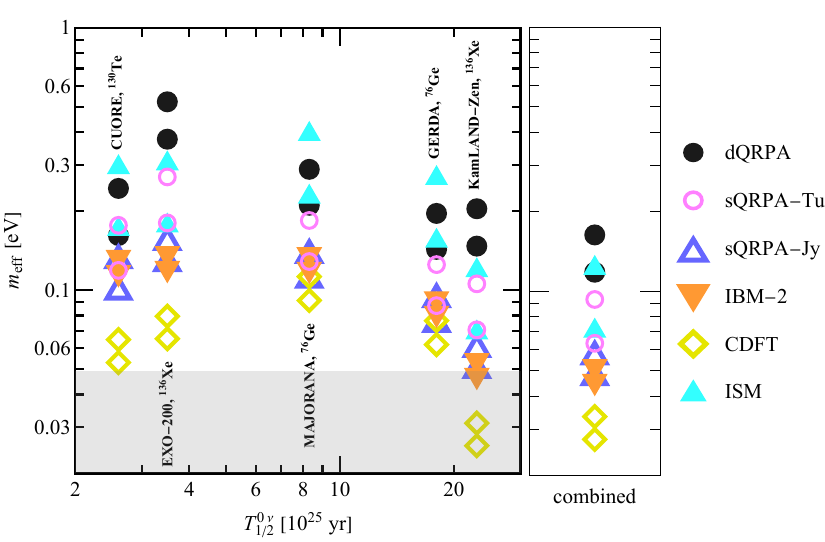}
	}
	\caption{90\% C.L. upper bounds of the $m_\mathrm{eff}$ for the various experiments.}
	\label{fig:meff-T}
\end{figure}

From Eq. \eqref{eq:half-life},
the NMEs will make an effect of the relationship of the $m_\mathrm{eff}$ ($m_\mathrm{eff}^{\nu}$) and the $T^{0\nu}_{1/2}$, and the values of NMEs are very different in various models, which contribute the dominated uncertainty to the constraints of the $T^{0\nu}_{1/2}$ as well as $m_\mathrm{eff}$.
Typically, Fig. \ref{fig:chi-meff} shows the $\Delta\chi^2$ as a function of $m_\mathrm{eff}$ in dQRPA model with $g_A = 1$ and Argonne src. 
We can see that the constraint of $m_\mathrm{eff}$ from GERDA is the most stringent, the constraint from KamLAND-Zen is competitive with that from GERDA, and constraint from EXO-200 is the weakest, although the constraint of $T^{0\nu}_{1/2}$  from EXO-200 is more stringent than CUORE, due to the different NME values and the phase spaces of different isotopes.
Note that the combined results are derived from $ \chi^2_{0\nu\beta\beta}$ by 
subtracting its minimum value regarding $m_{\rm eff}$.
More generally, we consider various NME models we discussed in last section. In this case,
things will be different due to the different prediction of the NMEs from different nuclear models.
Fig. \ref{fig:meff-T} shows the 90\% C.L. limits of $m_\mathrm{eff}$ for various experiments obtained by different nuclear models, and the results of combined analysis from $\chi^2_{0\nu\beta\beta}$ are also shown on the right side of the figure.
The different markers stand for the different nuclear models. 
The gray region represents the IMO range of $m_{\rm eff}^{\nu}$.
From Fig. \ref{fig:meff-T}, the dQRPA model will always get a weak limit, while the CDFT models will always get a strengthened limits due to their larger values for predicted NMEs.
We can regard the different NME calculation as a global NME uncertainty, and it will result in an order of magnitude uncertainty for the constraints on $m_\mathrm{eff}$. 
The most stringent limits are obtained from KamLAND-Zen experiment, and in the CDFT model, these constraints already exclude part of IMO region given the light neutrino exchange mechanism.
We can also find that the combined analysis of all the experiments will strength the constraints of $m_\mathrm{eff}$, but considering the NME uncertainty, it is still far away to reject IMO case. 

\begin{figure}[!htbp]
	\centering
	\subfigure{
		\includegraphics[width=0.8\textwidth]{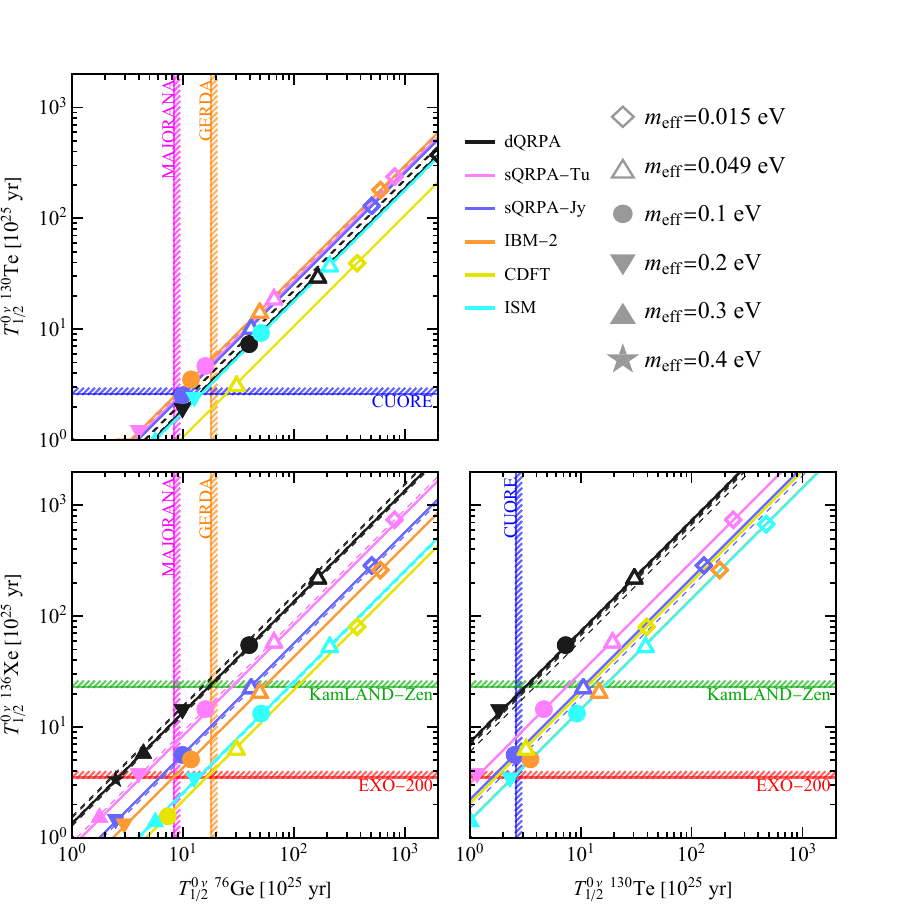}
	}
	\caption{Relation between the $T^{0\nu}_{1/2}$ in $^{76}$Ge, $^{136}$Xe and $^{130}$Te for different matrix element calculations, where the solide lines denote the NME in the case of $g_A=1.27$ and CD-bonn src while the dashed lines present other NME calculations.}
	\label{fig:fig_T_vs_T}
\end{figure}

The constraints of various experiments can be also shown in Fig. \ref{fig:fig_T_vs_T}, and the relation between the $T^{0\nu}_{1/2}$
of $^{76}\mathrm{Ge}$, $^{130}\mathrm{Te}$ and $^{136}\mathrm{Xe}$ for
different NME calculations is also shown.
The different markers stand for the different $m_{\rm eff}$ values.
According to Eq. \eqref{eq:half-life}, 
the half-lives between two different isotopes are linearly correlated on a logarithmic scale and the position of line is determined by the ratio of the phase spaces and the NME model.
For a certain pair of elements, the ratio of their phase spaces is a constant, so the slope of the line is determined by the ratio of NMEs.
In the pair of $^{76}\mathrm{Ge}$ and $^{136}\mathrm{Xe}$, the line of dQRPA is the uppermost line and the line of CDFT is in the bottom, which means that if the half-life of $^{76}\mathrm{Ge}$ is determined, the prediction of the half-life of $^{136}\mathrm{Xe}$ is longest in dQRPA calculation, and shortest in CDFT calculation. 
Considering the pair of $^{76}\mathrm{Ge}$ and $^{130}\mathrm{Te}$, when the half-life of $^{76}\mathrm{Ge}$ is determined, the half-life of $^{130}\mathrm{Te}$ is longer in sQRPA and IBM-2 calculation, and still shortest in CDFT calculation.
As for the pair of $^{130}\mathrm{Te}$ and $^{136}\mathrm{Xe}$, when the half-life of $^{130}\mathrm{Te}$ is determined, the half-life of $^{136}\mathrm{Xe}$ is longest in dQRPA calculation, and shortest in ISM calculation.
It is also easy to see that the relationship of the half-life between different elements in one NME model is almost independent of the assumption in this model, i.e. $g_A$ and srcs, although the half-life is different based on some certain value of $(m_\mathrm{eff})^2$ by different assumptions of the same NME model.  

According to Fig. \ref{fig:meff-T}, we can roughly get the corresponding constraints on the heavy neutrino parameters in the case of dQRPA model ($g_A=1$, Argonne src) from Figs. \ref{fig:meff-M1-Re1} and \ref{fig:meff-M1-M2} (mainly the red regions).
Next we will discuss the constraints on the minimal Type-I seesaw model depending on the NME models in more details.
In the SM, the 0$\nu\beta\beta$ decay experiments will constrain the three active neutrinos. And in the seesaw model, the RHNs will directly contribute to the $m_\mathrm{eff}$ of the 0$\nu\beta\beta$ decay, so the 0$\nu\beta\beta$ decay experiments constrain RHNs and three active neutrinos together.
The intrinsic relation in the seesaw mechanism leads to that only 4 free parameters of RHNs exist, and for convenience, we choose two masses of RHNs $M_1$, $M_2$, the absolute value of $R_{e1}^2$, and the mixing phase $\delta_{14}$.
The contribution of the active neutrino part from $m_{\rm eff}^{\nu}$ can be constrained by the neutrino oscillation measurement, cosmological observation, and $\beta$-decay experiments, but in the minimal Type-I seesaw model we considered in this work, the lightest neutrino mass is assumed to be zero, so only the oscillation measurement is relevant and considered in the analysis. To extract information involved in $m_{\rm eff}$, the final $\chi^2_{\rm tot}$ function is defined as
\begin{equation}
\begin{aligned}
    \chi^2_{\rm tot} &= \chi^2_{0\nu\beta\beta} + \chi^2_\mathrm{OSC}\;,
\end{aligned}
\label{chi2-tot}
\end{equation}
Here
\begin{equation}
  \chi^2_\mathrm{OSC} = \sum_{i} \frac{(\eta_i - \eta_i^{\rm BF})^2}{\sigma_i^2};, 
\end{equation}
where $\eta_i$ denote in turn the oscillation parameters $\{\theta_{12},~\theta_{13},~\Delta m_{21}^2,~\Delta m_{31}^2\}$ with $\Delta m_{ij}^2 \equiv m_i^2-m_j^2$ (for $ij=21,~31$), and $\eta_i^{\rm BF}$ and $\sigma_i$ stand for the best-fit values and symmetrized $1\sigma$ errors of $\eta_i$, respectively.

\begin{figure}[!htbp]
	\centering		\includegraphics[width=0.7\textwidth]{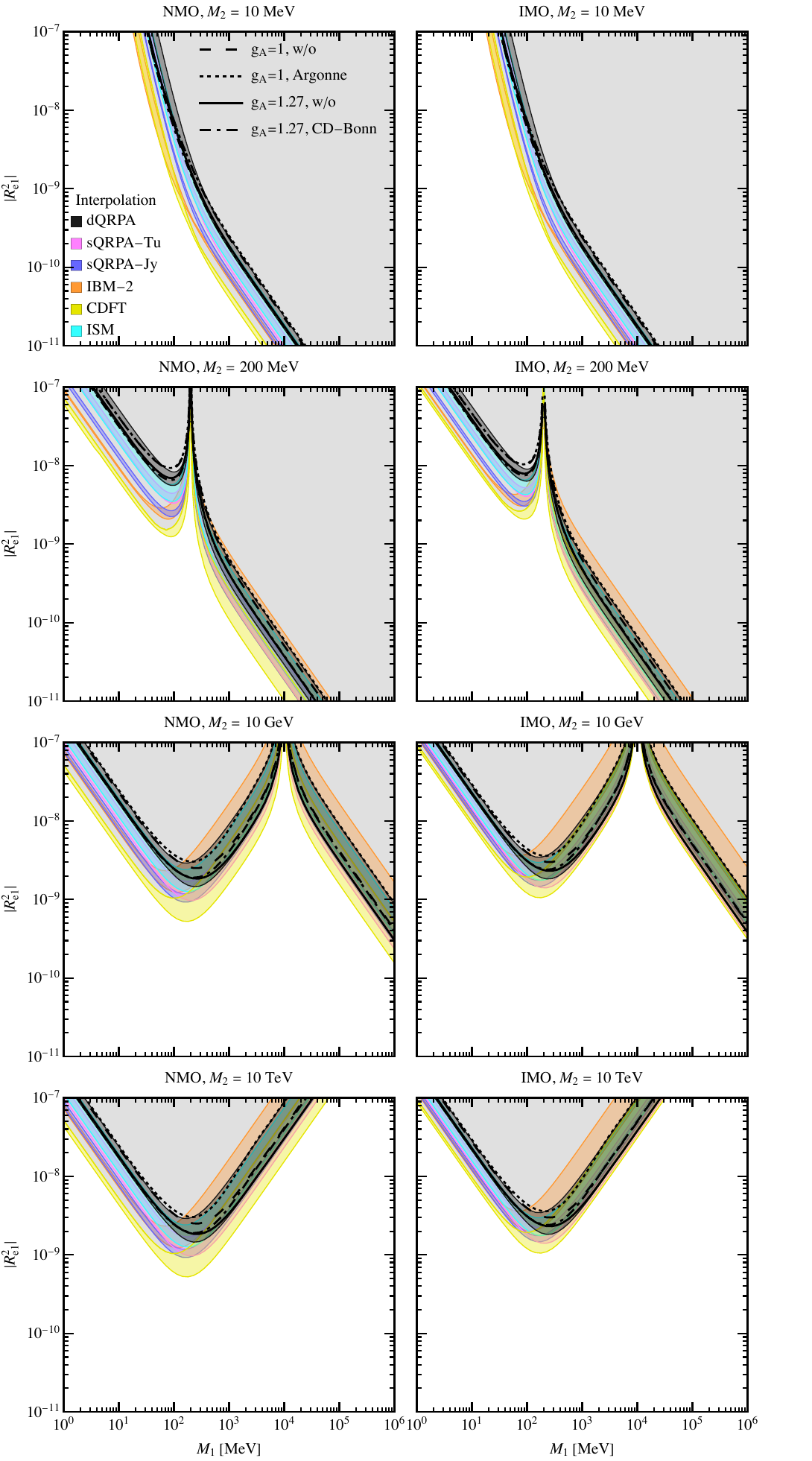}
	\caption{Excluded regions of the $M_1$ and $|R^2_\mathrm{e1}|$ given fixed $M_2$ in the case of different NME models by combining current results from  GERDA, MAJORANA, CUORE, KamLAND-Zen, and EXO-200 experiments.}
	\label{fig:cons-M1-Re1s}
\end{figure}

\begin{figure}[!htbp]
	\centering
		\subfigure{
		\includegraphics[width=0.7\textwidth]{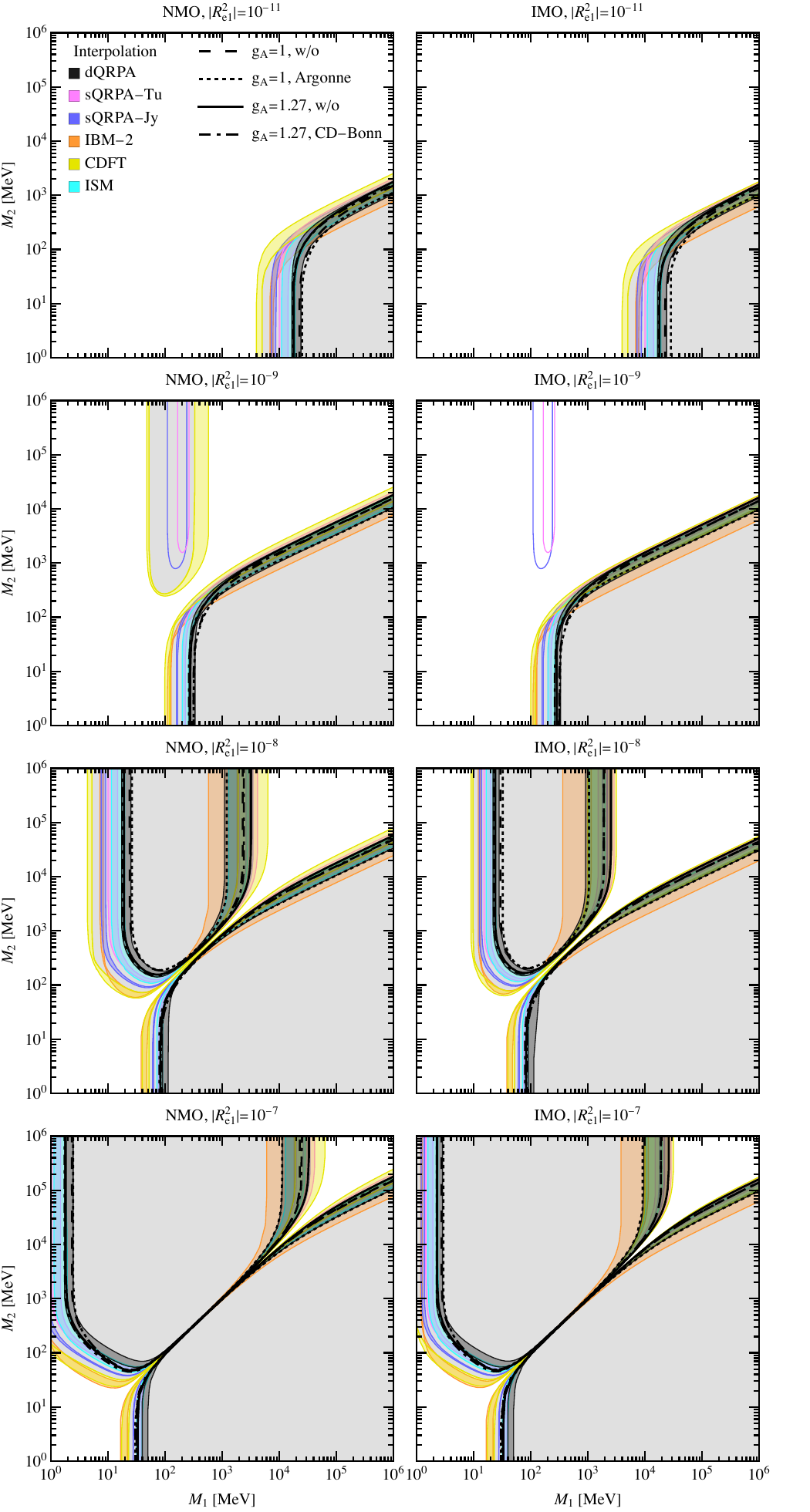}
	}
	\caption{Excluded regions of the $M_1$ and $M_2$ given fixed $|R_{e1}^2|$ in the case of different NME models by combining current results from  GERDA, MAJORANA, CUORE, KamLAND-Zen, and EXO-200 experiments.}
	\label{fig:cons-M1-M2}
\end{figure}

In Fig.~\ref{fig:cons-M1-Re1s}, we present the $3\sigma$ constraints of $M_1$ and $|R_{e1}^2|$ with the fixed values of $M_2=10~\mathrm{MeV}, 200~\mathrm{MeV}, 10~\mathrm{GeV}$, $10~\mathrm{TeV}$. The way is to first calculate the minima of $\chi^2_{\rm tot}$ by scanning the neutrino oscillation parameters in their $3\sigma$ ranges, Majorana phases in $[0,2\pi]$, and $\delta_{14}$ in $[0,\pi]$ for fixed values of $M_1$, $|R_{e1}^2|$ and $M_2$. Then the $3\sigma$ constraints
correspond to the contour $(\chi^2_{\rm tot} )_{\rm min}\simeq 11.83$ for two degrees of freedom.
Similarly, we present the $3\sigma$ constraints of $M_1$ and $M_2$ with the fixed values of $|R_{e1}^2|=10^{-11}, 10^{-9}, 10^{-8}, 10^{-7}$ in 
 Fig.~\ref{fig:cons-M1-M2}. Some discussions about these two figures are as follows.
 \begin{itemize}
     \item 

The NMO and IMO cases are similar while the former case will exclude a little more parameter space than the latter one. This is implied in Figs. \ref{fig:meff-M1-Re1} and \ref{fig:meff-M1-M2} when $m_{\rm eff}$ is not very small, for example the light-red region in the case of $\delta_{14}=\pi/2$ (namely, the three and four columns).
\item The gray regions represent the $3\sigma$ excluded regions derived from current experiments in the case of CDFT model. The colorful boundary bands display
the differences of NME models. The 
NME models can induce about one order uncertainties on the constraints of $M_1$, $M_2$ and $|R_{e1}^2|$ as shown in Figs. \ref{fig:cons-M1-Re1s} and \ref{fig:cons-M1-M2}.
\item The IBM-2 model can exclude more parameter space of RHNs, while the dQRPA model will exclude less  parameter space of RHNs. 
When $M_1$ is smaller than $200~\mathrm{MeV}$.
The reason is that the prediction of NMEs in light neutrino mass in the dQRPA model is smaller than the IBM-2 model, and in heavy neutrino mass the reverse applies.
For the similar reason, when $M_1$ is larger than $200~\mathrm{MeV}$, the case of ISM model is competitive with the dQRPA model.
\item The hierarchy of constraints changes at different $M_2$, since the hierarchy of the NMEs changes at different mass of RHNs. 
The variation of the upper bounds is small when $M_2$ is small, and is large when $M_2$ is large, especially in the case that $M_1$ is also large.
The variation in CDFT is the largest one, and the constraints in dQRPA and ISM models have the smallest variation. 
\item In the case of $M_2$ around $10~\mathrm{TeV}$, the constraints on two RHNs are similar to the case of one generation RHN.
The combined analysis in this case will constrain $|R_{e1}^2|<10^{-8}\sim 10^{-9}$ when $M_1$ is around $200~\mathrm{MeV}$, and $|R_{e1}^2|<10^{-8}\sim10^{-7}$ when $M_1$ is around $1~\mathrm{MeV}$.
In the case of small $M_2$, the combined analysis will exclude the large $M_1$ region, and with the $|R_{e1}^2|$ decreasing, the allowed region of $M_1$ is expanded.
\item In the case of $M_2$ is around $200~\mathrm{MeV}$ or $10~{\rm GeV}$, there is a peak also around $200~\mathrm{MeV}$ or $10~{\rm GeV}$. The constraints of $|R_{e1}^2|$ are much weaker compared with the constraints when $M_1$ deviated enough from $200~\mathrm{MeV}$ or $10~{\rm GeV}$. This is due to the degeneracy of two RHN masses, where the $m_{\rm eff}$ is not sensitive to $|R_{e1}^2|$ from Eq. \eqref{eq:meff}. 
\item From Fig.~\ref{fig:cons-M1-M2} showing
the $3\sigma$ excluded regions of $M_1$ and $M_2$ given fixed $|R_{e1}^2|$, it's obvious that as $|R_{e1}^2|$ is bigger,
more parameter space of $M_1$ and $M_2$
will be excluded. With $|R_{e1}^2|$ being very small, only the right corner can be excluded. In the case of $|R_{e1}^2|=10^{-9}$, the excluded regions for dQRPA, IBM-2 and ISM models consist only one part in the half plane of $M_1> M_2$. This is due to relatively small NME values leading to smaller upper limit of $m_{\rm eff}$ compared with sQRPA-Tu, sQRPA-Jy and CDFT models. 
\item In the lower two panels with $|R_{e1}^2|=10^{-8}$ and $|R_{e1}^2|=10^{-7}$ of Fig. \ref{fig:cons-M1-M2}, a big parameter space of $M_1$ and $M_2$ is excluded while the diagonal line $M_1=M_2$ still survives due to the small values of $m_{\rm eff}$ in this case. When both $M_1$ and $|R_{e1}^2|$ are small enough,
$M_2$ is allowed in whole mass range. This is because the third term in $m_{\rm eff}$, which is $|R_{e1}^2|e^{2 {\rm i} \delta_{14}} M_1 [f_{\beta}(M_1) - f_{\beta}(M_2) ]$, is suppressed according to Eq. \eqref{eq:meff}.
While in the case of $|R_{e1}^2|=10^{-7}$, only the lower left corner with small $M_1$ and $M_2$ and the upper right corner with big $M_1$ and $M_2$ are allowed, where the heavy neutrino mass mechanism and light neutrino mass mechanism dominate, respectively.
\end{itemize}

Other experiments have also studied the properties and limitations of right-handed neutrinos, in addition to neutrinoless double beta decay experiments. 
Cosmological observations, data from large hadron colliders, 
the neutrino probe~\cite{T2K:2019jwa,MicroBooNE:2019izn,MicroBooNE:2019izn} 
and charged lepton flavor violation search~\cite{Lindner:2016bgg} provide important information about the mass and interactions of right-handed neutrinos.
However, these results are often derived under the assumption of a single right-handed neutrino mixing with a single flavor, which may not be applicable in realistic seesaw models.
The constraints on the mixing $|R^2_{e1}|$ from the colliders  is the level of [$10^{-8}$, $10^{-9}$] by NA62~\cite{NA62:2020mcv} for $150\;{\rm MeV}<M_1<400\;{\rm MeV}$, and [$10^{-6}$, $10^{-7}$] by CHARM~\cite{CHARM:1985nku} for $400\;{\rm MeV}<M_1<2\;{\rm GeV}$, and the level of $10^{-4}$ for $M_1>2\;{\rm GeV}$ by Belle~\cite{Belle:2013ytx}, DELPHI~\cite{DELPHI:1996qcc}, ATLAS~\cite{ATLAS:2019kpx} and CMS~\cite{CMS:2018iaf}.
The results from the neutrino probe, T2K~\cite{T2K:2019jwa}, suggest the limit $|R^2_{e1}|$ varies from $10^{-7}$ to $10^{-9}$ for the mass range of $150\;{\rm MeV}<M_1<500\;{\rm MeV}$.
The case under the assumption of a single right-handed neutrino mixing with a single flavor is similar to the case of the one RHNs mass large enough, compared with the results of this work. We have showed that in this case,
the $0\nu\beta\beta$ experiment remains the strongest constraint on RHNs, particularly for heavier masses, as discussed in Ref. \cite{Faessler:2014kka}.
Although the constraints on the mixing matrix from the $0\nu\beta\beta$ experiment are weaker for lower masses, this range is not covered by other experiments. For masses around 1 GeV, the $0\nu\beta\beta$ experiment provides the strongest constraints on the mixing matrix elements, while other experiments provide similar or even better constraints, so that the results from these experiments can complement each other in this case.

\subsection{Sensitivities of future experiments}
For the sensitivities of future $0\nu\beta\beta$ experiments, we refer to the analysis scheme in Ref. \cite{Pompa:2023jxc} and use the $\chi^2$ function based on the Poisson distribution
\begin{equation}
\Delta\chi_{ij}^2 (m_{\rm eff}, (M_{0\nu})_{\alpha j};m_{\rm eff}^{\rm True},(M_{0\nu})_{\alpha i}^{\rm True})= 2\sum_{\alpha} (N_{\alpha j} - N_{\alpha i}^{\rm True} + N_{\alpha i}^{\rm True} {\rm In} \frac{ N_{\alpha i}^{\rm True}}{N_{\alpha j}})\;,
\end{equation}
where the subscript $\alpha$ denotes different isotopes/experiments and $i,j$ stand for different NME models.
$N_{\alpha i}^{\rm True} = B_{\alpha i} + S_{\alpha i} (m_{\rm eff}^{\rm True},(M_{0\nu})_{\alpha i}^{\rm True})$ is the total event number expected from a given experiment and some true NME value and true $m_{\rm eff}$ chosen by the Nature, and
$N_{\alpha j} = B_{\alpha j} + S_{\alpha j} (m_{\rm eff}^{},M_{\alpha j})$
represents the model to be compared with the ``true" one.  Note that the NME $M_{\alpha j}$ and $M_{\alpha i}^{\rm True}$ depend on neutrino mass, which differs from Ref. \cite{Pompa:2023jxc}. $S_{\alpha i}$ and $B_{\alpha i}$ are the expected numbers of signal and background events, respectively. They can be derived from $S_{\alpha i} (m_{\rm eff},~ M_{\alpha i}) = { \rm In} 2 \cdot N_A \cdot\varepsilon_{\alpha} \cdot(T_{1/2}^{0\nu})_{\alpha i}^{-1}\cdot T/(1~ {\rm yr})$, and $B_{\alpha} = b_{\alpha}\cdot \varepsilon_{\alpha}\cdot T/(1~ {\rm yr})$, 
with $N_A$ being the Avogadro's Number, $T$ the exposure time, $\varepsilon_{\alpha}$ the sensitive exposure in unit of ${\rm mol}\cdot{\rm yr}$,
$b_{\alpha}$ the background ratio in unit of ${\rm events}/({\rm mol}\cdot{\rm yr})$ and $(T_{1/2}^{0\nu})_{\alpha i}$ from Eq. \eqref{eq:half-life}. 

\begin{table}[h]
    \centering
    \begin{tabular}{c|c|c|c}
    \hline\hline
      Experiment  & Isotope & $\varepsilon$ [mol$\cdot$ yr] & $b$ [events/(mol$\cdot$ yr)] \\\hline
      LEGEND-1000   &$^{76}$Ge & 8736& 4.9 $\cdot 10^{-6}$ \\
     SuperNEMO   &$^{82}$Se & 185& 5.4 $\cdot 10^{-3}$ \\
     SNO+II   &$^{130}$Te & 8521& 5.7 $\cdot 10^{-3}$ \\
      nEXO  &$^{136}$Xe & 13700& 4.0 $\cdot 10^{-5}$ \\
         \hline\hline
    \end{tabular}
    \caption{The sensitive exposures and background ratios of $^{76}$Ge-based LEGEND-1000, $^{82}$Se-based SuperNEMO, $^{130}$Te based SNO+II and $^{136}$Xe-based nEXO \cite{Agostini:2022zub}.}
    \label{tab:exps-future}
\end{table}

Specially, ten year's exposure of
$^{76}$Ge-based LEGEND-1000 \cite{LEGEND:2021bnm}, $^{82}$Se-based SuperNEMO \cite{Kauer:2008em}, $^{130}$Te based SNO+II \cite{SNO:2021xpa} and $^{136}$Xe-based nEXO \cite{nEXO:2021ujk} are considered. The related background ratio and sensitive exposure are shown in Table \ref{tab:exps-future}, which we take from Table IV in Ref. \cite{Agostini:2022zub}.
For simplicity, we derive the $3\sigma$ sensitivities to type-I minimal seesaw parameters  by assuming that $m_{\rm eff}$ is zero, namely, observing no $0\nu\beta\beta$ signal. In this case, $N_{\alpha j}$ do not rely on NME 
model or the subscript $j$.
That is, the derived sensitivities do not depend on assuming which NME model is true. 

\begin{figure}
    \centering
    \includegraphics[width=0.7\textwidth]{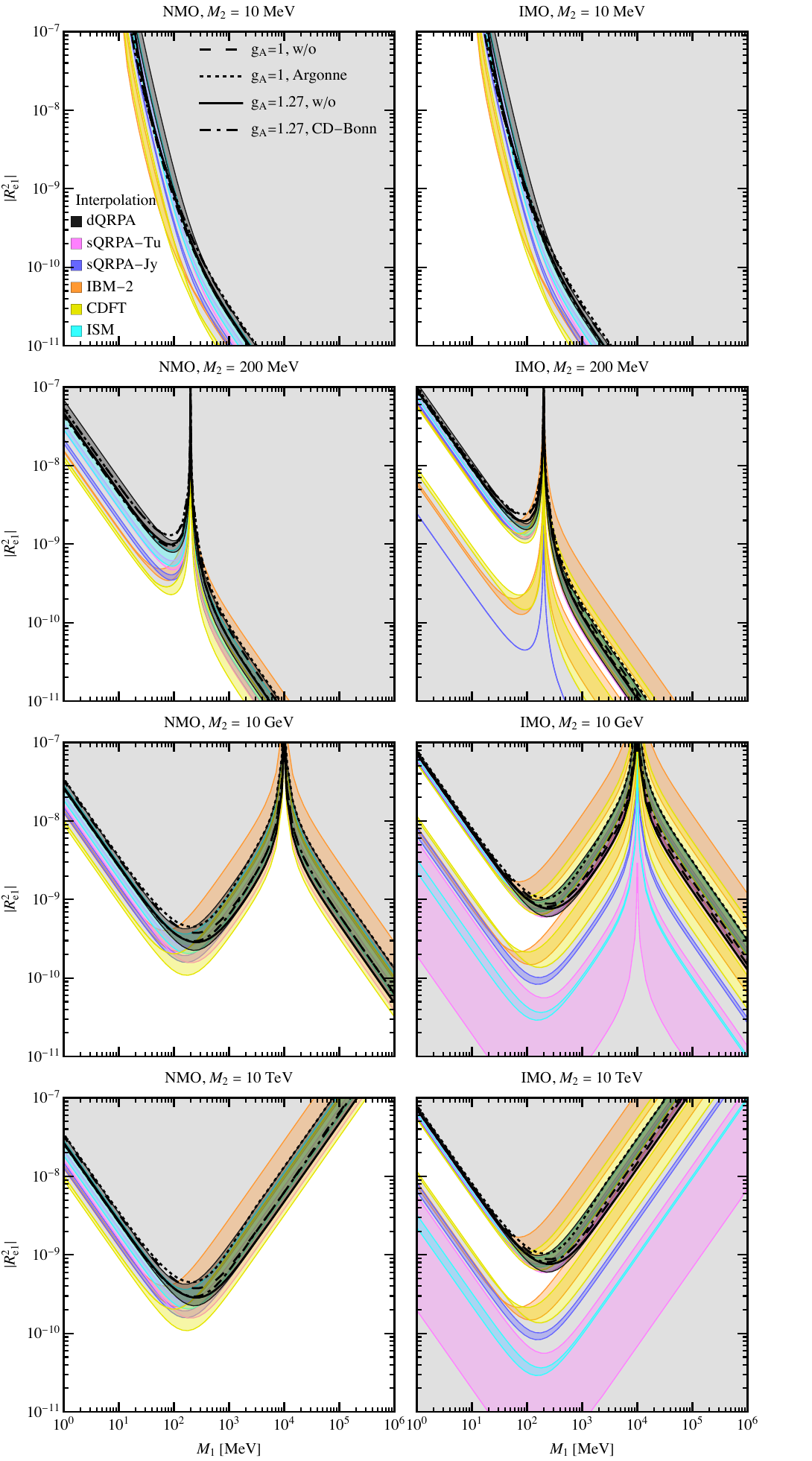}
    \caption{Sensitivity ranges of future $0\nu\beta\beta$ experiments to $M_1$ and $|R_{e1}^2|$ given fixed values of $M_2$ when we combine LEGEND-1000, SuperNEMO, SNO+II and nEXO by assuming $m_{\rm eff}^{\rm True}$= 0.}
    \label{fig:M1-Re1s-future}
\end{figure}

\begin{figure}
    \centering
    \includegraphics[width=0.7\textwidth]{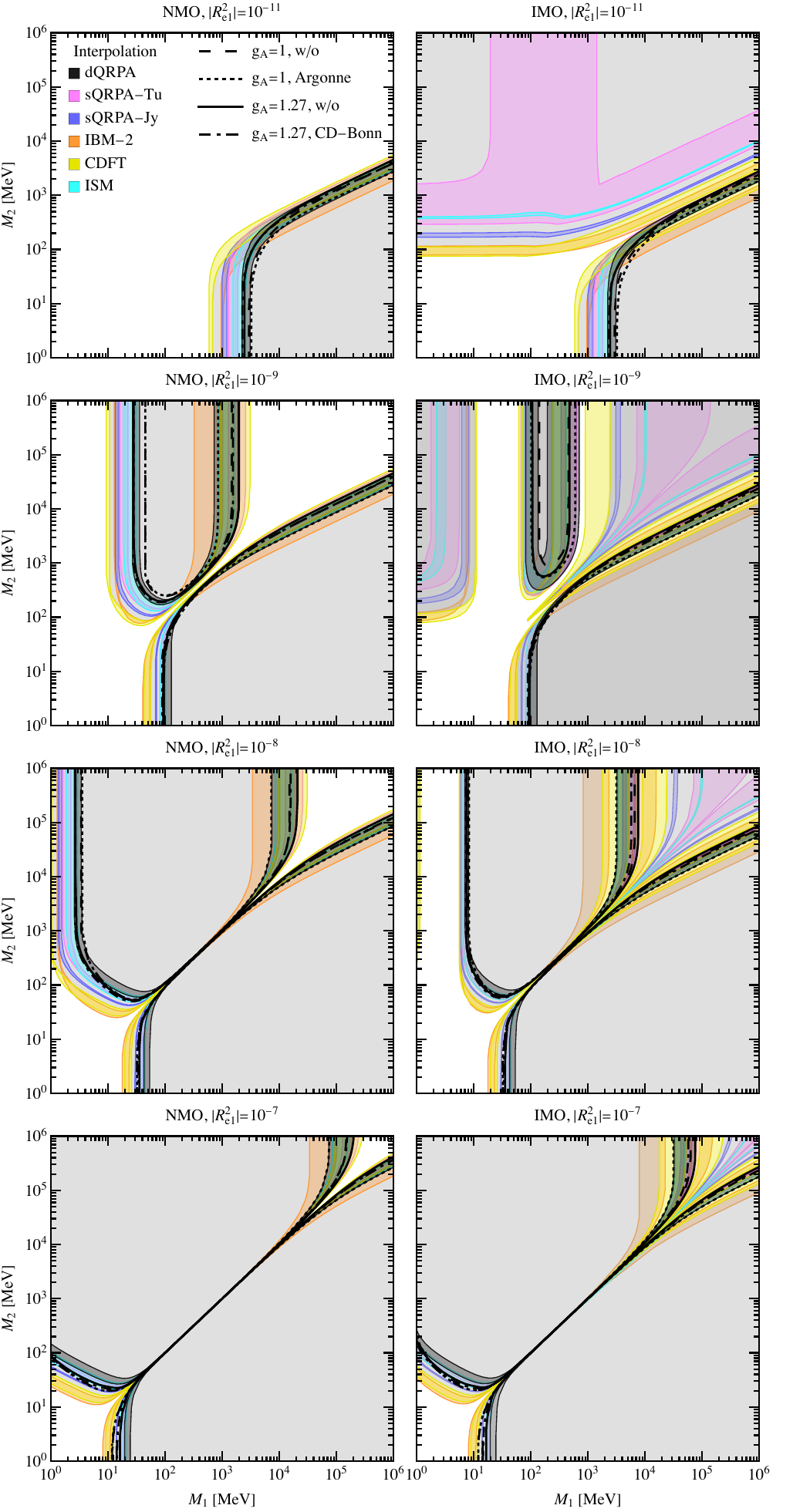}
    \caption{Sensitivity ranges of future experiments to $M_1$ and $M_2$ given fixed values of $|R_{e1}^2|$ when we combine LEGEND-1000, SuperNEMO, SNO+II and nEXO by assuming $m_{\rm eff}^{\rm True}$= 0.}
    \label{fig:M1-M2-future}
\end{figure}

Furthermore, we include the contribution of neutrino oscillation data and calculate 
the $3\sigma$ sensitivities to $M_1$ and $|R_{e1}^2|$ given fixed $M_2$ (or $M_1$ and $M_2$ given fixed $|R_{e1}^2|$) by considering $\chi_{\rm tot}^{\prime 2} =\Delta\chi_{ij}^2 (m_{\rm eff}, (M_{0\nu})_{\alpha j};m_{\rm eff}^{\rm True},(M_{0\nu})_{\alpha i}^{\rm True}) + \chi^2_{\rm OSC}$.  
By minimizing $\chi_{\rm tot}^{\prime 2}$ over the multi-parameter space of $\delta_{14}$, the related oscillation parameter in $m_{\rm eff}^{0\nu}$ as well as the Majorana phases and requiring $(\chi_{\rm tot}^{\prime 2})_{\rm min}=11.83$, we get the $3\sigma$ sensitive boundaries to $M_1$ and $|R_{e1}^2|$ (or $M_1$ and $M_2$) displayed in Fig.
\ref{fig:M1-Re1s-future} (Fig. \ref{fig:M1-M2-future}), where we consider both the NMO and the IMO cases. The gray region is the excluded region in the case of CDFT model and the colorful bands represent the excluded boundaries by considering other NME models. Note that in Figs. \ref{fig:M1-Re1s-future} and \ref{fig:M1-M2-future}, the bands are derived approximately by inputting the NME interpolation formula in Eq. \eqref{eq:NME-interpolation} from the light and heavy neutrino exchange mechanisms while the black lines display the results by  calculating mass-dependent NME in the case of dQRPA model.
The sensitivities of $M_1$, $M_2$ and $|R_{e1}^2|$ can be easily understood by combining the $m_{\rm eff}$ contours in Figs. \ref{fig:meff-M1-Re1} and \ref{fig:meff-M1-M2}, which imply the contours of some upper limit of $m_{\rm eff}$ derived from future experiments. The colorful bands roughly correspond to the sensitive contours of $m_{\rm eff}$ in a small range for each NME model if the contribution of oscillation data $\chi^2_{\rm OSC}$ is neglected.
The NMO and IMO case are very different and bigger parameter space will be excluded in the latter case.
Compared with the current limit, more parameter space is expected to be explored in the next-generation $0\nu\beta\beta$ experiments in the framework of minimal type-I seesaw model, especially in the IMO case.
This is because that we assume $m_{\rm eff}^{\rm True}=0$, meaning that big range of the IMO case is excluded at $3\sigma$ level. 
Note that the big pink band in the upper right panel of Fig. \ref{fig:M1-M2-future} corresponds to four cases of sQRPA-Tu in Table \ref{tab:LNE} and Table \ref{tab:HNE}, namely, $g_A=1.27$ and Argonne src, $g_A=1.27$ and CD-Bonn src, $g_A=1$ and Argonne src, as well as $g_A=1$ and CD-Bonn src. The second case leads to the lower boundary of the band and the first and fourth cases also result in boundaries close to the lower boundary. While in the third case, we get the upper boundary of the band mostly due to very different value of $\delta_{14}$ to be taken when performing the minimization of $\chi_{\rm tot}^{\prime 2}$ over the multi-parameter space including $\delta_{14}$ to derive the boundary. From Fig. \ref{fig:meff-M1-M2}, we know that the contours of $m_{\rm eff}$ can be very sensitive to $\delta_{14}$.

The above discussion is a conservative treatment with the hypothesis that no positive signal is observed. On the other hand, if we take bigger value of $m_{\rm eff}^{\rm True}$,
such as $50~{\rm meV}$, meaning that most future $0\nu\beta\beta$ experiments are expected to observe some positive signals, the sensitive ranges of $M_1$, $M_2$ and $|R_{e1}^2|$ will depend significantly on the assumed true NME model. And it is even possible to discriminate among different NME models given some good benchmarks. It is also expected that a large portion of the parameter space can be exclude at $3\sigma$ by referring to Figs. \ref{fig:meff-M1-Re1} and \ref{fig:meff-M1-M2}.
Note that the LEGEND-1000, SuperNEMO, SNO+II and nEXO 
taken above are all competitive proposals for each of the isotopes in consideration. For other future experiments, similar calculation can be performed.

\section{Conclusion} \label{sec:conclusion}
In this work, we have discussed the Majorana neutrino mass dependent nuclear matrix element of the neutrinoless double beta decay and its implications for the minimal Type-I seesaw model. We have compared the explicit many-body calculations and naive extrapolations of the mass dependent NME and discussed their significant differences. We have derived limits on the parameter space of this model from the current $0\nu\beta\beta$ data, and compared with other experiments in a limiting case. Our results show that the constraints from $0\nu\beta\beta$ experiments is still the strongest one, and the NME plays a crucial role in determining the sensitivity of $0\nu\beta\beta$ experiments to the Majorana neutrino masses and mixing elements.
Furthermore, the sensitive reach of future $0\nu\beta\beta$ proposals is also studied depending on various NME models. More parameter space of the minimal type-I seesaw is expected to be explored.
Our work highlights the importance of accurate NME calculations for interpreting the experimental data and constraining new physics models. Further studies are needed to improve our understanding of NMEs and their uncertainties, as well as to explore other models beyond the minimal Type-I seesaw model.

\section{Acknowledgements}
Y.F.~Li is grateful to Profs.~Ke Han and Gaosong Li for helpful discussions on the CUORE and EXO-200 results respectively. The work of Y.F.~Li and Y.Y.~Zhang is supported by National Natural Science Foundation of China under Grant No.~12075255, ~12075254 and No.~11835013. Y.Y.~Zhang is also supported by China Postdoctoral Science Foundation funded project.
D.L.~Fang is supported by CAS from the ``Light of West" program and the ``from zero to one" program and by MOST from the National Key Research and Development Program of China (2021YFA1601300).

\bibliographystyle{JHEP_improved}
\bibliography{ref}
	
\end{document}